\documentclass[aps,prx,10pt,twocolumn,notitlepage,superscriptaddress]{revtex4-2}

\usepackage{booktabs}
\usepackage{tikz}
\usepackage{quantikz}
\usepackage[breaklinks=true]{hyperref}

\begin{document}

\title{Solving MNIST with a globally trained Mixture of Quantum Experts}

\author{Paolo Alessandro Xavier Tognini}  \email{paolo.tognini@sns.it}
\affiliation{ Scuola Normale Superiore, Piazza dei Cavalieri 7, 56126 Pisa, Italy} 

\author{Leonardo Banchi} \email{leonardo.banchi@unifi.it}
\affiliation{Department of Physics and Astronomy, University of Florence,
via G. Sansone 1, I-50019 Sesto Fiorentino (FI), Italy}
\affiliation{ INFN Sezione di Firenze, via G. Sansone 1, I-50019, Sesto Fiorentino (FI), Italy }

\author{Giacomo De Palma} \email{giacomo.depalma@unibo.it}
\affiliation{University of Bologna, Department of Mathematics, Piazza di Porta San Donato 5, 40126 Bologna BO, Italy}

\begin{abstract}
  We propose a new quantum neural network for image classification, which is able to classify the parity of the MNIST dataset with full resolution with a test accuracy of up to $97.5\%$ without any classical pre-processing or post-processing.
  Our architecture is based on a mixture of experts whose model function is the sum of the model functions of each expert.
  We encode the input with amplitude encoding, which allows us to encode full-resolution MNIST images with $10$ qubits and to implement a convolution on the whole image with just a single one-qubit gate.
  Our training algorithm is based on training all the experts together, which significantly improves trainability with respect to training each expert independently.
  In fact, in the limit of infinitely many experts, our training algorithm can perfectly fit the training data. Our results demonstrate the potential of our quantum neural network to achieve high-accuracy image classification with minimal quantum resources, paving the way for more scalable and efficient quantum machine learning models.
\end{abstract}

\maketitle

\section{Introduction}
As the NISQ era \cite{Preskill_2018} thrives with ever greater quantum computers, and as we are on the verge of an Artificial Intelligence revolution, interest ignites in a relatively recent discipline called Quantum Machine Learning \cite{Biamonte_2017,Cerezo_2022} (or QML), which explores the use of quantum computation to enhance the capabilities of machine learning algorithms.
Nonetheless, the quest for quantum advantage in QML is ongoing, without clear and unequivocal success \cite{AaronsonHHL,schuld2022quantum}. There are two main approaches: the first one tries to use results in complexity theory to prove rigorous advantages in certain settings, which typically require a rather unconventional access to the data. 
An example is the HHL algorithm for linear systems of equations \cite{HHL,Gily_n_2019,cifuentes2024quantumcomputationalcomplexitymatrix}.
The second approach is heuristic and uses Variational Quantum Algorithms \cite{VQA}, i.e., parametric quantum circuits trained by gradient descent. For their similarities with classical Neural Networks, variational quantum algorithms are also called quantum neural networks.
Despite the excitement, quantum neural networks can suffer from several problems:
\begin{itemize}
	\item barren plateaus \cite{BarrenPlateaus,Cerezo_2021,napp2022quantifyingbarrenplateauphenomenon}: the magnitude of the gradients decreases exponentially in the number of layers, which means that for deep architectures, the landscape of the loss function is almost flat (a plateau) and is extremely difficult to navigate;
	\item there is an expressibility-trainability trade-off \cite{ExpressibilityBarrenPlateaus}, analogous to the classical bias-variance trade-off;
    \item the cost function typically contains many local minima, so a poor initialization of the parameters can make the model untrainable even without the presence of barren plateaus \cite{Anschuetz_2022}.
\end{itemize}

Although the barren plateau phenomenon has been extensively studied in the literature and some mitigation strategies exist \cite{grant2019initialization,rudolph2023synergistic,liu2023mitigating}, little is known about the other issues of trainability. In classical computing, the advance of deep learning has shown the potential of overparameterized neural networks with billions of parameters \cite{lecun2015deep}. Intuitively, more parameters can allow the model to circumvent local optima. The empirical success of deep learning is also due to the backpropagation algorithm, which allows efficient training \cite{lillicrap2020backpropagation}. However, getting backpropagation scaling in variational quantum circuits is complicated \cite{abbas2023quantum,gilyen2019optimizing} due to peculiar quantum features, such as no-cloning. In this work, we introduce a new architecture, based on a mixture of experts, which avoids the complications of quantum backpropagation while still improving trainability.

We focus on the task of vision, i.e. the classification of images, and in particular on the classification of the MNIST dataset of handwritten digits (see Appendix \ref{sec:appendixMNIST}).
In quantum vision, two lines of research have been developed. The first consists of hybrid classical-quantum algorithms, where either the input is classically pre-processed before being encoded into the quantum circuit \cite{qViT,qSlowFeatureAnalysis}, or the output of the quantum circuit is classically post-processed \cite{QuantumKitchenSinks,PostVariationalQNNs}. In both cases, the actual contribution of the quantum part of the algorithm is hard to distinguish from the contribution of the classical part.
The second line of research consists of purely quantum algorithms.
As the $28\times28$ resolution of MNIST images is too high for basis encoding, where the intensity of each pixel is encoded in a different qubit, the previous quantum algorithms for MNIST usually resorted to extremely downsampled images, with a resolution reduced even to $4 \times 4$ \cite{GoogleQuantumMNIST}.

\begin{figure}[h!]
	\centering
	\includegraphics[width=1.0\linewidth]{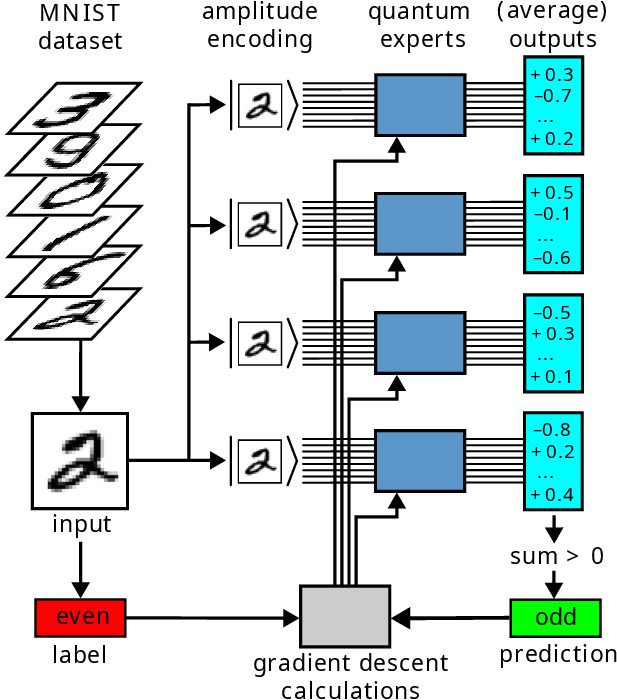}
	\caption{Scheme of our Mixture of Quantum Experts (MoQE) architecture. Each expert encodes an input MNIST image in $10$ qubits through amplitude encoding, applies a parametric quantum circuit (blue box) and measures each qubit in the $Z$ basis; their average values are thus numbers in the interval $[-1,+1]$. The whole procedure is repeated $n$ times, where $n$ is the number of experts ($n=4$ in the figure), with different parameters. The sign of the sum of the outputs of all experts gives the binary prediction, which is confronted with the true label. Through (classical) gradient descent calculations, the parameters of the quantum experts are updated together.}
	\label{fig:SchemeQuantumCircuit}
\end{figure}

In this paper, we propose a \emph{purely quantum architecture} that can potentially solve (at least for binary classification) the task of Vision for images of hundreds or thousands of pixels (as in the MNIST dataset) while using only $10$ to $20$ qubits thanks to amplitude encoding, which encodes the coefficients of the input vector in the amplitudes of the state of the qubits.

\section{Our architecture}
In classical ML, a standard architecture that solves Vision is the Convolutional Neural Network (CNN) (see Appendix~\ref{sec:appendixCNN}), which looks for features at increasingly larger scales via a combination of convolutional and pooling layers.
Our quantum circuit architecture is based on the same idea: to look for features on small scales first and then to use the results to look for features on larger scales.
A quantum generalization of CNNs has been proposed in \cite{qCNN,qCNN2}. The main difference between such quantum CNNs and our architecture is that, thanks to amplitude encoding, in our architecture each convolution operation requires only a single quantum gate. In fact, with amplitude encoding, a single $1$-qubit gate performs a convolution, as is shown in Appendix \ref{sec:1 qubit convolution}).

Our scalable Mixture of Quantum Experts (MoQE) architecture, 
which is schematically shown in \autoref{fig:SchemeQuantumCircuit}, 
differs from others in the literature \cite{qSlowFeatureAnalysis,GoogleQuantumMNIST,QuantumKitchenSinks,PostVariationalQNNs} in some key aspects:
\begin{enumerate}
  \item the classification is done purely at the quantum level, without any classical neural network pre-processing or post-processing; 
  \item we leverage amplitude encoding to process full-resolution images without downsampling;
  \item inspired by CNNs, the variational circuit is defined assuming correlations between neighboring pixels;
  \item we use a mixture of experts, where each expert is a quantum neural network and, crucially, all experts are trained together. 
\end{enumerate}

As a relevant problem, we focus on the classification of the parity of MNIST digits, which has been widely used in the Quantum Vision literature \cite{qSlowFeatureAnalysis,GoogleQuantumMNIST,QuantumKitchenSinks,PostVariationalQNNs}. 
MNIST images have a resolution of $28\times28$ pixels.
We encode each full-resolution image in the amplitudes of the wavefunction of $10$ qubits as follows.
The $28\times28$ image is embedded in a $32\times32$ image by padding white pixels at the borders.
For any $\mathbf x,\,\mathbf y=0,\,\ldots,\,31$, let $c_{\mathbf x,\mathbf y}$ be the intensity of the pixel in the range $[0,1]$.
Let us express $\mathbf x=(x_0,\,x_1,\,\ldots,\,x_4) $ and $\mathbf y=(y_0,\,y_1,\,\ldots,\,y_4)$ in binary using 5 bits each, where the first bit encodes the most significant digit of the coordinate, and so on. 
Then, by mapping each bit into a qubit, the image is encoded in the $10$-qubit wavefunction
\begin{equation}
|\psi\rangle = \frac{1}{\mathcal{N}}\sum_{\mathbf x,\,\mathbf y\in [0,1]^5} c_{\mathbf x,\mathbf y}|x_0,y_0,\dots, x_4,y_4\rangle\,,
\label{eq:imagestate}
\end{equation}
where $\mathcal{N}$ is the factor that normalizes the wavefunction, which is different for each image.
Amplitude encoding can be compiled in a quantum circuit in several ways: for instance, by first finding the corresponding matrix product state representation \cite{latorre2005image} and then mapping that into a quantum circuit \cite{cramer2010efficient}.

\begin{figure*}[!th]
    \begin{center}
        \includegraphics[width=0.95\textwidth]{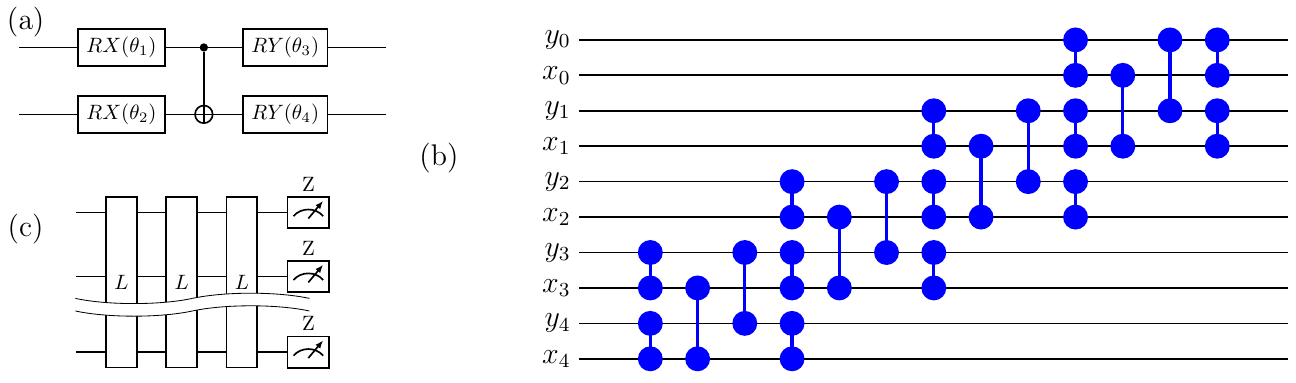}
    \end{center}
	\caption{
            (a) Our elementary $2$-qubit gate with $4$ parameters. This will be the building block of the layers of our quantum circuit.
            (b) The architecture of each layer: information goes from lower resolution pixel (below) to higher resolution ones (above), moving upward like climbing a ladder. Each blue gate is the $2$-qubit gate presented in (a).
            (c) The final $10$-qubit quantum circuit contains three repetitions of the above layer (denoted with $L$), with the $Z$ measurements at the end -- for clarity, only three qubits are shown.
    }
	\label{fig:gates}
\end{figure*}

We are now ready to build our circuit from the bottom up.
The building block of our architecture is the $2$-qubit gate with four parameters of \autoref{fig:gates}(a).
Such gates are used to build the layer in \autoref{fig:gates}(b) in which we look for patterns at small scales and then use the output to look for patterns at higher and higher scales. In a layer, the $2$-qubit gates iteratively act on couples of qubits in \eqref{eq:imagestate} of the form $(x_i,y_i)$ or $(x_i,x_{i+1})$ or $(y_i,y_{i+1})$. We employ $3$ layers,
for a total of $252$ parameters, and then 
each qubit is measured in the Pauli-$Z$ basis -- 
see \autoref{fig:gates}(c).
The output is given by the normalized sum of the outcomes. The normalization is such that the empirical variance of the result on the MNIST images is equal to one.

Our architecture enjoys a very useful feature: if the qubits are arranged on a line as in \autoref{fig:gates}, the $2$-qubit gates of the main circuit only act on qubits with distance not greater than two. Moreover, if we arrange the $x_i$ and $y_i$ qubits on two adjacent lines forming a $2\times5$ grid, then all $2$-qubit gates only act on neighboring qubits. Therefore, a small quantum computer with a $2 \times 5$ grid of qubits capable of single-qubit Pauli rotations and CNOT gates between neighboring qubits can efficiently implement our architecture (except for amplitude encoding).

\subsection{Mixture of (Quantum) Experts}
To increase the number of parameters without increasing either the number of qubits or the number of layers, we opted for a variant of a technique called Mixture of Experts (MoE), widely used in classical AI \cite{MoEreview}, even in large language models such as DeepSeek-V3 \cite{liu2024deepseek}. 
Our quantum neural network architecture, schematically shown in \autoref{fig:SchemeQuantumCircuit}, 
is composed of $n$ experts, where each expert consists of the quantum circuit described in \autoref{fig:gates} and has the same input image, but different independent parameters. 
The output of the whole network is the sum of the outputs of each expert divided by a normalization constant proportional to $\sqrt{n}$, and the binary classification prediction is given by the sign of the total output. The normalization factor is irrelevant for the classification task, but affects the cost.

Training consists of minimizing the square loss cost function $C = (f - \ell)^2$,
where $f$ is the output of the circuit for a given image and $\ell\in\{\pm1\}$ is the true label, i.e. the parity of the digit represented in the image.
We train our network with gradient descent,
where we employ the parameter-shift rule \cite{ParameterShiftRule} to compute the cost gradient.
We stress that we do not train each expert independently, but rather train the global output of the network as a whole.
Therefore, the gradient descent update of each parameter will generally also depend on the parameters of the other experts.

The mixture of experts will be hard to simulate classically whenever the single experts are hard to simulate.
The number of parameters increases linearly with the number of experts and so does the computational cost.
Nevertheless, we will empirically show that, over a certain regime, the overall computational cost to achieve a certain accuracy is lower than that of training fewer experts for more epochs.

\subsection{The limit of infinitely many experts}
The idea of an architecture made by a mixture of quantum experts has been inspired by recent results on the infinite-width limit of quantum neural networks \cite{Girardi_2025,hernandez2024quantitativeconvergencetrainedquantum,Abedi_2023}.
Such results state that in the limit of infinite width (i.e. of infinitely many qubits), the probability distribution of the model function of a quantum neural network trained with gradient descent on a supervised learning problem converges to a Gaussian process, provided that the past light cones of the measured qubits are small enough and that the network is not affected by barren plateaus.
The mean and the covariance of the limit Gaussian process can be computed analytically, and reveal that the cost function decays exponentially in the training time and therefore the training is always able to perfectly fit the training examples in a time that grows at most polynomially in the number of qubits \cite[Theorems 5.4 and 5.9]{Girardi_2025}.
In particular, if barren plateaus are not present, the issue of bad local minima disappears in the limit of infinite width.
Furthermore, the model function remains smooth in the limit despite the divergence in the number of parameters.

These results can be directly applied to our quantum neural network, since our mixture of experts can be considered as a single network whose width is the sum of the widths of each expert. The condition on small light cones is automatically satisfied as the experts are disconnected.
For our network, the limit of infinite width becomes the limit of infinitely many experts.
In such a limit, the results above guarantee that the network is trainable and that the model function remains smooth.

\section{Experiments}
We used all images from the MNIST dataset and performed simulations with 
$n = 1, 2, 4, 8, 16, 32$ experts, respectively.
The code was written with Pennylane \cite{bergholm2018pennylane}, using the Pytorch backend. We employed the widely used Adam algorithm \cite{Adam} for gradient descent.
The cost function for each gradient descent step was computed on a batch of size $4$.

\begin{figure}[t!]
	\centering
	\includegraphics[width=1.0\linewidth]{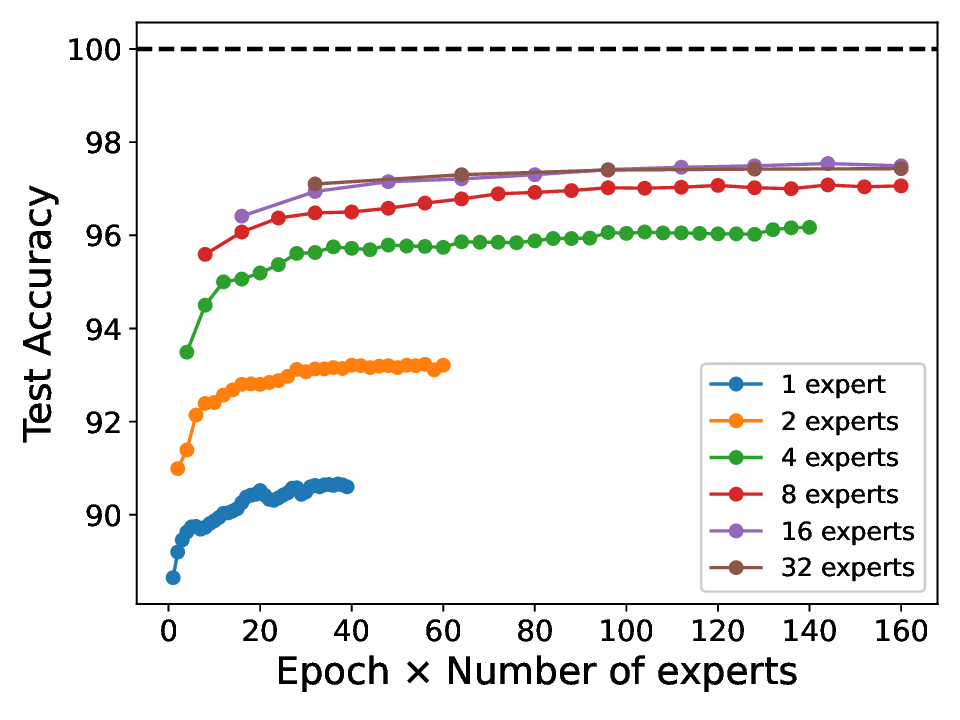}
	\caption{Test accuracy in function of the computation, which in our model is proportional to the number of training epochs multiplied by the number of experts. This plot allows us to see what is the best way to allocate the compute to reach the maximum accuracy. We have cut some of the data to ease the comparison between different numbers of experts.}
	\label{fig:test vs epochs rescaled}
\end{figure}

A summary of the results of our simulations is given in \autoref{fig:test vs epochs rescaled} and \autoref{tab:table1}, while 
extended numerical results are presented in Appendix~\ref{sec:full}. 
The horizontal axis in \autoref{fig:test vs epochs rescaled} shows the total computational cost, which is 
proportional to the number of epochs times the number of experts, since each expert requires 
evaluation and measurement of the same quantum circuit, but with different parameters. 
In spite of the (linearly) higher computational cost required to train more experts, we show that the test accuracy for the same computational cost is actually higher when the number of experts trained together is increased, until saturation at about $16$ to $32$ 
experts. 
This is the central result of our paper and shows the remarkable advantage of globally training a mixture of quantum experts. 

\begin{table}[t]
	\centering
	\begin{tabular}{l c c r}
	    \toprule
	    \#experts & train acc. & test acc. & \#epochs \\
	    \midrule
	    1  & 90.66\% &90.60\% & 39 \\
	    2  & 93.37\% &93.21\% & 30 \\
	    4  & 96.18\% &96.17\% & 35 \\
	    8  & 96.86\% &97.06\% & 20 \\
	    16 & 97.38\% &\textbf{97.54}\% & 12 \\
	    32 & \textbf{97.55}\% &97.43\% &  5 \\
	    \bottomrule
	\end{tabular}
    \caption{Results for the simulation of our model with different number of experts. We tabulate the training and test accuracy at the end of the training. We highlighted in bold the best result of each column.}
    \label{tab:table1}
\end{table}

The maximum training accuracy saturates at $97.5\%$ for about $16$ experts.
While the obtained accuracy is still below what is possible to achieve with classical neural networks,
even when using a comparable number of parameters (see Appendix~\ref{sec:appendixCNN}), we believe that the obtained accuracy is close to the optimal one allowed by our model class and that the better performances of other classical models are due to their higher or different non-linearity. 
In fact, due to amplitude encoding, ignoring the normalization factor, 
the function class defined by our current MoQE is quadratic in the input, since the expectation value of the output of each expert has the form $\bra \psi \hat O \ket \psi$ for some observable $\hat O$, and since $\ket \psi$ is linear in the input when using amplitude encoding -- we can indeed ignore the normalization constant since it is the same for all experts and only provides a global factor that does not affect the sign and hence the model prediction.
In Appendix~\ref{sec:AppendixQuadratic} we show the results obtained by training a general classical quadratic model, 
which basically reaches $98\%$ test accuracy, only slightly better than our best performing MoQE model. 
However, such a general quadratic classifier requires $\approx 3\cdot 10^5$ parameters, 
significantly more than those used in our largest MoQE. Therefore, we believe that our architecture approaches the best performing classical model within its function class, yet with fewer parameters,
and that higher accuracy can only be obtained by changing the function class. 
One possible way to do so in future studies is to mix the amplitude encoding with data re-uploading layers \cite{perez2020data}, which are capable of enhancing the model's non-linearity.

\section{Conclusions}
In this paper, we have proposed a scalable Mixture of Quantum Experts (MoQE) architecture for quantum Vision inspired by the CNN and MoE architectures of classical Machine Learning.
Our MoQE architecture has been validated by $97.5\%$ test accuracy in parity prediction in the MNIST dataset. Thanks to amplitude encoding, our quantum neural network needs few qubits ($10$ for the MNIST images) and can perform the equivalent of an entire classical convolutional layer with a single $1$-qubit quantum gate.
Our architecture is purely quantum and does not require any classical pre-processing or post-processing.
The mixture of experts allows our architecture to increase the number of parameters without increasing the number of required qubits.
The model function is not bounded to be quadratic, since the input wavefunction of each expert can consist of multiple copies of the image.
Although state-of-the-art classical CNNs and humans can reach an accuracy of $99.8\%$ \cite{MaxAccuracyMNIST}, this is the first time that a purely quantum architecture is successfully applied to the non-coarse grained MNIST database.

\section*{Acknowledgements}
GDP has been supported by the HPC Italian National Centre for HPC, Big Data and Quantum Computing -- Proposal code CN00000013 -- CUP J33C22001170001 and by the Italian Extended Partnership PE01 -- FAIR Future Artificial Intelligence Research -- Proposal code PE00000013 -- CUP J33C22002830006 under the MUR National Recovery and Resilience Plan funded by the European Union -- NextGenerationEU.
GDP and LB have been supported by the European Union -- NextGenerationEU under the National Recovery and Resilience Plan (PNRR) -- Mission 4 Education and research -- Component 2 From research to business -- Investment 1.1 Notice Prin 2022 -- DD N. 104 del 2/2/2022, from title ``understanding the LEarning process of QUantum Neural networks (LeQun)'', proposal code 2022WHZ5XH -- CUP J53D23003890006.
LB has been supported the European Union’s Horizon Europe research and innovation program under EPIQUE Project GA No.~101135288. 
GDP is a member of the ``Gruppo Nazionale per la Fisica Matematica (GNFM)'' of the ``Istituto Nazionale di Alta Matematica ``Francesco Severi'' (INdAM)''.

\bibliographystyle{apsrev4-1}
\setlength{\emergencystretch}{10pt}
\bibliography{q_references}

\begin{thebibliography}{43}%
\makeatletter
\providecommand \@ifxundefined [1]{%
 \@ifx{#1\undefined}
}%
\providecommand \@ifnum [1]{%
 \ifnum #1\expandafter \@firstoftwo
 \else \expandafter \@secondoftwo
 \fi
}%
\providecommand \@ifx [1]{%
 \ifx #1\expandafter \@firstoftwo
 \else \expandafter \@secondoftwo
 \fi
}%
\providecommand \natexlab [1]{#1}%
\providecommand \enquote  [1]{``#1''}%
\providecommand \bibnamefont  [1]{#1}%
\providecommand \bibfnamefont [1]{#1}%
\providecommand \citenamefont [1]{#1}%
\providecommand \href@noop [0]{\@secondoftwo}%
\providecommand \href [0]{\begingroup \@sanitize@url \@href}%
\providecommand \@href[1]{\@@startlink{#1}\@@href}%
\providecommand \@@href[1]{\endgroup#1\@@endlink}%
\providecommand \@sanitize@url [0]{\catcode `\\12\catcode `\$12\catcode
  `\&12\catcode `\#12\catcode `\^12\catcode `\_12\catcode `\%12\relax}%
\providecommand \@@startlink[1]{}%
\providecommand \@@endlink[0]{}%
\providecommand \url  [0]{\begingroup\@sanitize@url \@url }%
\providecommand \@url [1]{\endgroup\@href {#1}{\urlprefix }}%
\providecommand \urlprefix  [0]{URL }%
\providecommand \Eprint [0]{\href }%
\providecommand \doibase [0]{http://dx.doi.org/}%
\providecommand \selectlanguage [0]{\@gobble}%
\providecommand \bibinfo  [0]{\@secondoftwo}%
\providecommand \bibfield  [0]{\@secondoftwo}%
\providecommand \translation [1]{[#1]}%
\providecommand \BibitemOpen [0]{}%
\providecommand \bibitemStop [0]{}%
\providecommand \bibitemNoStop [0]{.\EOS\space}%
\providecommand \EOS [0]{\spacefactor3000\relax}%
\providecommand \BibitemShut  [1]{\csname bibitem#1\endcsname}%
\let\auto@bib@innerbib\@empty
\bibitem [{\citenamefont {Preskill}(2018)}]{Preskill_2018}%
  \BibitemOpen
  \bibfield  {author} {\bibinfo {author} {\bibfnamefont {J.}~\bibnamefont
  {Preskill}},\ }\href {\doibase 10.22331/q-2018-08-06-79} {\bibfield
  {journal} {\bibinfo  {journal} {Quantum}\ }\textbf {\bibinfo {volume} {2}},\
  \bibinfo {pages} {79} (\bibinfo {year} {2018})}\BibitemShut {NoStop}%
\bibitem [{\citenamefont {Biamonte}\ \emph {et~al.}(2017)\citenamefont
  {Biamonte}, \citenamefont {Wittek}, \citenamefont {Pancotti}, \citenamefont
  {Rebentrost}, \citenamefont {Wiebe},\ and\ \citenamefont
  {Lloyd}}]{Biamonte_2017}%
  \BibitemOpen
  \bibfield  {author} {\bibinfo {author} {\bibfnamefont {J.}~\bibnamefont
  {Biamonte}}, \bibinfo {author} {\bibfnamefont {P.}~\bibnamefont {Wittek}},
  \bibinfo {author} {\bibfnamefont {N.}~\bibnamefont {Pancotti}}, \bibinfo
  {author} {\bibfnamefont {P.}~\bibnamefont {Rebentrost}}, \bibinfo {author}
  {\bibfnamefont {N.}~\bibnamefont {Wiebe}}, \ and\ \bibinfo {author}
  {\bibfnamefont {S.}~\bibnamefont {Lloyd}},\ }\href {\doibase
  10.1038/nature23474} {\bibfield  {journal} {\bibinfo  {journal} {Nature}\
  }\textbf {\bibinfo {volume} {549}},\ \bibinfo {pages} {195–202} (\bibinfo
  {year} {2017})}\BibitemShut {NoStop}%
\bibitem [{\citenamefont {Cerezo}\ \emph {et~al.}(2022)\citenamefont {Cerezo},
  \citenamefont {Verdon}, \citenamefont {Huang}, \citenamefont {Cincio},\ and\
  \citenamefont {Coles}}]{Cerezo_2022}%
  \BibitemOpen
  \bibfield  {author} {\bibinfo {author} {\bibfnamefont {M.}~\bibnamefont
  {Cerezo}}, \bibinfo {author} {\bibfnamefont {G.}~\bibnamefont {Verdon}},
  \bibinfo {author} {\bibfnamefont {H.-Y.}\ \bibnamefont {Huang}}, \bibinfo
  {author} {\bibfnamefont {L.}~\bibnamefont {Cincio}}, \ and\ \bibinfo {author}
  {\bibfnamefont {P.~J.}\ \bibnamefont {Coles}},\ }\href {\doibase
  10.1038/s43588-022-00311-3} {\bibfield  {journal} {\bibinfo  {journal}
  {Nature Computational Science}\ }\textbf {\bibinfo {volume} {2}},\ \bibinfo
  {pages} {567–576} (\bibinfo {year} {2022})}\BibitemShut {NoStop}%
\bibitem [{\citenamefont {Aaronson}(2015)}]{AaronsonHHL}%
  \BibitemOpen
  \bibfield  {author} {\bibinfo {author} {\bibfnamefont {S.}~\bibnamefont
  {Aaronson}},\ }\href {\doibase 10.1038/nphys3272} {\bibfield  {journal}
  {\bibinfo  {journal} {Nature Phys.}\ }\textbf {\bibinfo {volume} {11}},\
  \bibinfo {pages} {291} (\bibinfo {year} {2015})}\BibitemShut {NoStop}%
\bibitem [{\citenamefont {Schuld}\ and\ \citenamefont
  {Killoran}(2022)}]{schuld2022quantum}%
  \BibitemOpen
  \bibfield  {author} {\bibinfo {author} {\bibfnamefont {M.}~\bibnamefont
  {Schuld}}\ and\ \bibinfo {author} {\bibfnamefont {N.}~\bibnamefont
  {Killoran}},\ }\href@noop {} {\bibfield  {journal} {\bibinfo  {journal} {Prx
  Quantum}\ }\textbf {\bibinfo {volume} {3}},\ \bibinfo {pages} {030101}
  (\bibinfo {year} {2022})}\BibitemShut {NoStop}%
\bibitem [{\citenamefont {Harrow}\ \emph {et~al.}(2009)\citenamefont {Harrow},
  \citenamefont {Hassidim},\ and\ \citenamefont {Lloyd}}]{HHL}%
  \BibitemOpen
  \bibfield  {author} {\bibinfo {author} {\bibfnamefont {A.~W.}\ \bibnamefont
  {Harrow}}, \bibinfo {author} {\bibfnamefont {A.}~\bibnamefont {Hassidim}}, \
  and\ \bibinfo {author} {\bibfnamefont {S.}~\bibnamefont {Lloyd}},\ }\href
  {\doibase 10.1103/physrevlett.103.150502} {\bibfield  {journal} {\bibinfo
  {journal} {Physical Review Letters}\ }\textbf {\bibinfo {volume} {103}}
  (\bibinfo {year} {2009}),\ 10.1103/physrevlett.103.150502}\BibitemShut
  {NoStop}%
\bibitem [{\citenamefont {Gilyén}\ \emph {et~al.}(2019)\citenamefont
  {Gilyén}, \citenamefont {Su}, \citenamefont {Low},\ and\ \citenamefont
  {Wiebe}}]{Gily_n_2019}%
  \BibitemOpen
  \bibfield  {author} {\bibinfo {author} {\bibfnamefont {A.}~\bibnamefont
  {Gilyén}}, \bibinfo {author} {\bibfnamefont {Y.}~\bibnamefont {Su}},
  \bibinfo {author} {\bibfnamefont {G.~H.}\ \bibnamefont {Low}}, \ and\
  \bibinfo {author} {\bibfnamefont {N.}~\bibnamefont {Wiebe}},\ }in\ \href
  {\doibase 10.1145/3313276.3316366} {\emph {\bibinfo {booktitle} {Proceedings
  of the 51st Annual ACM SIGACT Symposium on Theory of Computing}}},\ \bibinfo
  {series and number} {STOC ’19}\ (\bibinfo  {publisher} {ACM},\ \bibinfo
  {year} {2019})\ p.\ \bibinfo {pages} {193–204}\BibitemShut {NoStop}%
\bibitem [{\citenamefont {Cifuentes}\ \emph {et~al.}(2024)\citenamefont
  {Cifuentes}, \citenamefont {Wang}, \citenamefont {Silva}, \citenamefont
  {Berta},\ and\ \citenamefont
  {Aolita}}]{cifuentes2024quantumcomputationalcomplexitymatrix}%
  \BibitemOpen
  \bibfield  {author} {\bibinfo {author} {\bibfnamefont {S.}~\bibnamefont
  {Cifuentes}}, \bibinfo {author} {\bibfnamefont {S.}~\bibnamefont {Wang}},
  \bibinfo {author} {\bibfnamefont {T.~L.}\ \bibnamefont {Silva}}, \bibinfo
  {author} {\bibfnamefont {M.}~\bibnamefont {Berta}}, \ and\ \bibinfo {author}
  {\bibfnamefont {L.}~\bibnamefont {Aolita}},\ }\href
  {https://arxiv.org/abs/2410.13937} {\enquote {\bibinfo {title} {Quantum
  computational complexity of matrix functions},}\ } (\bibinfo {year} {2024}),\
  \Eprint {http://arxiv.org/abs/2410.13937} {arXiv:2410.13937 [quant-ph]}
  \BibitemShut {NoStop}%
\bibitem [{\citenamefont {Cerezo}\ \emph
  {et~al.}(2021{\natexlab{a}})\citenamefont {Cerezo}, \citenamefont
  {Arrasmith}, \citenamefont {Babbush}, \citenamefont {Benjamin}, \citenamefont
  {Endo}, \citenamefont {Fujii}, \citenamefont {McClean}, \citenamefont
  {Mitarai}, \citenamefont {Yuan}, \citenamefont {Cincio},\ and\ \citenamefont
  {Coles}}]{VQA}%
  \BibitemOpen
  \bibfield  {author} {\bibinfo {author} {\bibfnamefont {M.}~\bibnamefont
  {Cerezo}}, \bibinfo {author} {\bibfnamefont {A.}~\bibnamefont {Arrasmith}},
  \bibinfo {author} {\bibfnamefont {R.}~\bibnamefont {Babbush}}, \bibinfo
  {author} {\bibfnamefont {S.~C.}\ \bibnamefont {Benjamin}}, \bibinfo {author}
  {\bibfnamefont {S.}~\bibnamefont {Endo}}, \bibinfo {author} {\bibfnamefont
  {K.}~\bibnamefont {Fujii}}, \bibinfo {author} {\bibfnamefont {J.~R.}\
  \bibnamefont {McClean}}, \bibinfo {author} {\bibfnamefont {K.}~\bibnamefont
  {Mitarai}}, \bibinfo {author} {\bibfnamefont {X.}~\bibnamefont {Yuan}},
  \bibinfo {author} {\bibfnamefont {L.}~\bibnamefont {Cincio}}, \ and\ \bibinfo
  {author} {\bibfnamefont {P.~J.}\ \bibnamefont {Coles}},\ }\href {\doibase
  10.1038/s42254-021-00348-9} {\bibfield  {journal} {\bibinfo  {journal}
  {Nature Reviews Physics}\ }\textbf {\bibinfo {volume} {3}},\ \bibinfo {pages}
  {625–644} (\bibinfo {year} {2021}{\natexlab{a}})}\BibitemShut {NoStop}%
\bibitem [{\citenamefont {McClean}\ \emph {et~al.}(2018)\citenamefont
  {McClean}, \citenamefont {Boixo}, \citenamefont {Smelyanskiy}, \citenamefont
  {Babbush},\ and\ \citenamefont {Neven}}]{BarrenPlateaus}%
  \BibitemOpen
  \bibfield  {author} {\bibinfo {author} {\bibfnamefont {J.~R.}\ \bibnamefont
  {McClean}}, \bibinfo {author} {\bibfnamefont {S.}~\bibnamefont {Boixo}},
  \bibinfo {author} {\bibfnamefont {V.~N.}\ \bibnamefont {Smelyanskiy}},
  \bibinfo {author} {\bibfnamefont {R.}~\bibnamefont {Babbush}}, \ and\
  \bibinfo {author} {\bibfnamefont {H.}~\bibnamefont {Neven}},\ }\href
  {\doibase 10.1038/s41467-018-07090-4} {\bibfield  {journal} {\bibinfo
  {journal} {Nature Communications}\ }\textbf {\bibinfo {volume} {9}} (\bibinfo
  {year} {2018}),\ 10.1038/s41467-018-07090-4}\BibitemShut {NoStop}%
\bibitem [{\citenamefont {Cerezo}\ \emph
  {et~al.}(2021{\natexlab{b}})\citenamefont {Cerezo}, \citenamefont {Sone},
  \citenamefont {Volkoff}, \citenamefont {Cincio},\ and\ \citenamefont
  {Coles}}]{Cerezo_2021}%
  \BibitemOpen
  \bibfield  {author} {\bibinfo {author} {\bibfnamefont {M.}~\bibnamefont
  {Cerezo}}, \bibinfo {author} {\bibfnamefont {A.}~\bibnamefont {Sone}},
  \bibinfo {author} {\bibfnamefont {T.}~\bibnamefont {Volkoff}}, \bibinfo
  {author} {\bibfnamefont {L.}~\bibnamefont {Cincio}}, \ and\ \bibinfo {author}
  {\bibfnamefont {P.~J.}\ \bibnamefont {Coles}},\ }\href {\doibase
  10.1038/s41467-021-21728-w} {\bibfield  {journal} {\bibinfo  {journal}
  {Nature Communications}\ }\textbf {\bibinfo {volume} {12}} (\bibinfo {year}
  {2021}{\natexlab{b}}),\ 10.1038/s41467-021-21728-w}\BibitemShut {NoStop}%
\bibitem [{\citenamefont
  {Napp}(2022)}]{napp2022quantifyingbarrenplateauphenomenon}%
  \BibitemOpen
  \bibfield  {author} {\bibinfo {author} {\bibfnamefont {J.}~\bibnamefont
  {Napp}},\ }\href {https://arxiv.org/abs/2203.06174} {\enquote {\bibinfo
  {title} {Quantifying the barren plateau phenomenon for a model of
  unstructured variational ans\"{a}tze},}\ } (\bibinfo {year} {2022}),\ \Eprint
  {http://arxiv.org/abs/2203.06174} {arXiv:2203.06174 [quant-ph]} \BibitemShut
  {NoStop}%
\bibitem [{\citenamefont {Holmes}\ \emph {et~al.}(2022)\citenamefont {Holmes},
  \citenamefont {Sharma}, \citenamefont {Cerezo},\ and\ \citenamefont
  {Coles}}]{ExpressibilityBarrenPlateaus}%
  \BibitemOpen
  \bibfield  {author} {\bibinfo {author} {\bibfnamefont {Z.}~\bibnamefont
  {Holmes}}, \bibinfo {author} {\bibfnamefont {K.}~\bibnamefont {Sharma}},
  \bibinfo {author} {\bibfnamefont {M.}~\bibnamefont {Cerezo}}, \ and\ \bibinfo
  {author} {\bibfnamefont {P.~J.}\ \bibnamefont {Coles}},\ }\href {\doibase
  10.1103/prxquantum.3.010313} {\bibfield  {journal} {\bibinfo  {journal} {PRX
  Quantum}\ }\textbf {\bibinfo {volume} {3}} (\bibinfo {year} {2022}),\
  10.1103/prxquantum.3.010313}\BibitemShut {NoStop}%
\bibitem [{\citenamefont {Anschuetz}\ and\ \citenamefont
  {Kiani}(2022)}]{Anschuetz_2022}%
  \BibitemOpen
  \bibfield  {author} {\bibinfo {author} {\bibfnamefont {E.~R.}\ \bibnamefont
  {Anschuetz}}\ and\ \bibinfo {author} {\bibfnamefont {B.~T.}\ \bibnamefont
  {Kiani}},\ }\href {\doibase 10.1038/s41467-022-35364-5} {\bibfield  {journal}
  {\bibinfo  {journal} {Nature Communications}\ }\textbf {\bibinfo {volume}
  {13}} (\bibinfo {year} {2022}),\ 10.1038/s41467-022-35364-5}\BibitemShut
  {NoStop}%
\bibitem [{\citenamefont {Grant}\ \emph {et~al.}(2019)\citenamefont {Grant},
  \citenamefont {Wossnig}, \citenamefont {Ostaszewski},\ and\ \citenamefont
  {Benedetti}}]{grant2019initialization}%
  \BibitemOpen
  \bibfield  {author} {\bibinfo {author} {\bibfnamefont {E.}~\bibnamefont
  {Grant}}, \bibinfo {author} {\bibfnamefont {L.}~\bibnamefont {Wossnig}},
  \bibinfo {author} {\bibfnamefont {M.}~\bibnamefont {Ostaszewski}}, \ and\
  \bibinfo {author} {\bibfnamefont {M.}~\bibnamefont {Benedetti}},\ }\href@noop
  {} {\bibfield  {journal} {\bibinfo  {journal} {Quantum}\ }\textbf {\bibinfo
  {volume} {3}},\ \bibinfo {pages} {214} (\bibinfo {year} {2019})}\BibitemShut
  {NoStop}%
\bibitem [{\citenamefont {Rudolph}\ \emph {et~al.}(2023)\citenamefont
  {Rudolph}, \citenamefont {Miller}, \citenamefont {Motlagh}, \citenamefont
  {Chen}, \citenamefont {Acharya},\ and\ \citenamefont
  {Perdomo-Ortiz}}]{rudolph2023synergistic}%
  \BibitemOpen
  \bibfield  {author} {\bibinfo {author} {\bibfnamefont {M.~S.}\ \bibnamefont
  {Rudolph}}, \bibinfo {author} {\bibfnamefont {J.}~\bibnamefont {Miller}},
  \bibinfo {author} {\bibfnamefont {D.}~\bibnamefont {Motlagh}}, \bibinfo
  {author} {\bibfnamefont {J.}~\bibnamefont {Chen}}, \bibinfo {author}
  {\bibfnamefont {A.}~\bibnamefont {Acharya}}, \ and\ \bibinfo {author}
  {\bibfnamefont {A.}~\bibnamefont {Perdomo-Ortiz}},\ }\href {\doibase
  10.1038/s41467-023-43908-6} {\bibfield  {journal} {\bibinfo  {journal}
  {Nature Communications}\ }\textbf {\bibinfo {volume} {14}} (\bibinfo {year}
  {2023}),\ 10.1038/s41467-023-43908-6}\BibitemShut {NoStop}%
\bibitem [{\citenamefont {Liu}\ \emph {et~al.}(2023)\citenamefont {Liu},
  \citenamefont {Sun}, \citenamefont {Wu}, \citenamefont {Han},\ and\
  \citenamefont {Guo}}]{liu2023mitigating}%
  \BibitemOpen
  \bibfield  {author} {\bibinfo {author} {\bibfnamefont {H.-Y.}\ \bibnamefont
  {Liu}}, \bibinfo {author} {\bibfnamefont {T.-P.}\ \bibnamefont {Sun}},
  \bibinfo {author} {\bibfnamefont {Y.-C.}\ \bibnamefont {Wu}}, \bibinfo
  {author} {\bibfnamefont {Y.-J.}\ \bibnamefont {Han}}, \ and\ \bibinfo
  {author} {\bibfnamefont {G.-P.}\ \bibnamefont {Guo}},\ }\href@noop {}
  {\bibfield  {journal} {\bibinfo  {journal} {New Journal of Physics}\ }\textbf
  {\bibinfo {volume} {25}},\ \bibinfo {pages} {013039} (\bibinfo {year}
  {2023})}\BibitemShut {NoStop}%
\bibitem [{\citenamefont {LeCun}\ \emph {et~al.}(2015)\citenamefont {LeCun},
  \citenamefont {Bengio},\ and\ \citenamefont {Hinton}}]{lecun2015deep}%
  \BibitemOpen
  \bibfield  {author} {\bibinfo {author} {\bibfnamefont {Y.}~\bibnamefont
  {LeCun}}, \bibinfo {author} {\bibfnamefont {Y.}~\bibnamefont {Bengio}}, \
  and\ \bibinfo {author} {\bibfnamefont {G.}~\bibnamefont {Hinton}},\
  }\href@noop {} {\bibfield  {journal} {\bibinfo  {journal} {nature}\ }\textbf
  {\bibinfo {volume} {521}},\ \bibinfo {pages} {436} (\bibinfo {year}
  {2015})}\BibitemShut {NoStop}%
\bibitem [{\citenamefont {Lillicrap}\ \emph {et~al.}(2020)\citenamefont
  {Lillicrap}, \citenamefont {Santoro}, \citenamefont {Marris}, \citenamefont
  {Akerman},\ and\ \citenamefont {Hinton}}]{lillicrap2020backpropagation}%
  \BibitemOpen
  \bibfield  {author} {\bibinfo {author} {\bibfnamefont {T.~P.}\ \bibnamefont
  {Lillicrap}}, \bibinfo {author} {\bibfnamefont {A.}~\bibnamefont {Santoro}},
  \bibinfo {author} {\bibfnamefont {L.}~\bibnamefont {Marris}}, \bibinfo
  {author} {\bibfnamefont {C.~J.}\ \bibnamefont {Akerman}}, \ and\ \bibinfo
  {author} {\bibfnamefont {G.}~\bibnamefont {Hinton}},\ }\href@noop {}
  {\bibfield  {journal} {\bibinfo  {journal} {Nature Reviews Neuroscience}\
  }\textbf {\bibinfo {volume} {21}},\ \bibinfo {pages} {335} (\bibinfo {year}
  {2020})}\BibitemShut {NoStop}%
\bibitem [{\citenamefont {Abbas}\ \emph {et~al.}(2023)\citenamefont {Abbas},
  \citenamefont {King}, \citenamefont {Huang}, \citenamefont {Huggins},
  \citenamefont {Movassagh}, \citenamefont {Gilboa},\ and\ \citenamefont
  {McClean}}]{abbas2023quantum}%
  \BibitemOpen
  \bibfield  {author} {\bibinfo {author} {\bibfnamefont {A.}~\bibnamefont
  {Abbas}}, \bibinfo {author} {\bibfnamefont {R.}~\bibnamefont {King}},
  \bibinfo {author} {\bibfnamefont {H.-Y.}\ \bibnamefont {Huang}}, \bibinfo
  {author} {\bibfnamefont {W.~J.}\ \bibnamefont {Huggins}}, \bibinfo {author}
  {\bibfnamefont {R.}~\bibnamefont {Movassagh}}, \bibinfo {author}
  {\bibfnamefont {D.}~\bibnamefont {Gilboa}}, \ and\ \bibinfo {author}
  {\bibfnamefont {J.}~\bibnamefont {McClean}},\ }\href@noop {} {\bibfield
  {journal} {\bibinfo  {journal} {Advances in Neural Information Processing
  Systems}\ }\textbf {\bibinfo {volume} {36}},\ \bibinfo {pages} {44792}
  (\bibinfo {year} {2023})}\BibitemShut {NoStop}%
\bibitem [{\citenamefont {Gily{\'e}n}\ \emph {et~al.}(2019)\citenamefont
  {Gily{\'e}n}, \citenamefont {Arunachalam},\ and\ \citenamefont
  {Wiebe}}]{gilyen2019optimizing}%
  \BibitemOpen
  \bibfield  {author} {\bibinfo {author} {\bibfnamefont {A.}~\bibnamefont
  {Gily{\'e}n}}, \bibinfo {author} {\bibfnamefont {S.}~\bibnamefont
  {Arunachalam}}, \ and\ \bibinfo {author} {\bibfnamefont {N.}~\bibnamefont
  {Wiebe}},\ }in\ \href@noop {} {\emph {\bibinfo {booktitle} {Proceedings of
  the Thirtieth Annual ACM-SIAM Symposium on Discrete Algorithms}}}\ (\bibinfo
  {organization} {SIAM},\ \bibinfo {year} {2019})\ pp.\ \bibinfo {pages}
  {1425--1444}\BibitemShut {NoStop}%
\bibitem [{\citenamefont {Cherrat}\ \emph {et~al.}(2024)\citenamefont
  {Cherrat}, \citenamefont {Kerenidis}, \citenamefont {Mathur}, \citenamefont
  {Landman}, \citenamefont {Strahm},\ and\ \citenamefont {Li}}]{qViT}%
  \BibitemOpen
  \bibfield  {author} {\bibinfo {author} {\bibfnamefont {E.~A.}\ \bibnamefont
  {Cherrat}}, \bibinfo {author} {\bibfnamefont {I.}~\bibnamefont {Kerenidis}},
  \bibinfo {author} {\bibfnamefont {N.}~\bibnamefont {Mathur}}, \bibinfo
  {author} {\bibfnamefont {J.}~\bibnamefont {Landman}}, \bibinfo {author}
  {\bibfnamefont {M.}~\bibnamefont {Strahm}}, \ and\ \bibinfo {author}
  {\bibfnamefont {Y.~Y.}\ \bibnamefont {Li}},\ }\href {\doibase
  10.22331/q-2024-02-22-1265} {\bibfield  {journal} {\bibinfo  {journal}
  {Quantum}\ }\textbf {\bibinfo {volume} {8}},\ \bibinfo {pages} {1265}
  (\bibinfo {year} {2024})}\BibitemShut {NoStop}%
\bibitem [{\citenamefont {Kerenidis}\ and\ \citenamefont
  {Luongo}(2020)}]{qSlowFeatureAnalysis}%
  \BibitemOpen
  \bibfield  {author} {\bibinfo {author} {\bibfnamefont {I.}~\bibnamefont
  {Kerenidis}}\ and\ \bibinfo {author} {\bibfnamefont {A.}~\bibnamefont
  {Luongo}},\ }\href {\doibase 10.1103/physreva.101.062327} {\bibfield
  {journal} {\bibinfo  {journal} {Physical Review A}\ }\textbf {\bibinfo
  {volume} {101}} (\bibinfo {year} {2020}),\
  10.1103/physreva.101.062327}\BibitemShut {NoStop}%
\bibitem [{\citenamefont {Wilson}\ \emph {et~al.}(2019)\citenamefont {Wilson},
  \citenamefont {Otterbach}, \citenamefont {Tezak}, \citenamefont {Smith},
  \citenamefont {Polloreno}, \citenamefont {Karalekas}, \citenamefont {Heidel},
  \citenamefont {Alam}, \citenamefont {Crooks},\ and\ \citenamefont
  {da~Silva}}]{QuantumKitchenSinks}%
  \BibitemOpen
  \bibfield  {author} {\bibinfo {author} {\bibfnamefont {C.~M.}\ \bibnamefont
  {Wilson}}, \bibinfo {author} {\bibfnamefont {J.~S.}\ \bibnamefont
  {Otterbach}}, \bibinfo {author} {\bibfnamefont {N.}~\bibnamefont {Tezak}},
  \bibinfo {author} {\bibfnamefont {R.~S.}\ \bibnamefont {Smith}}, \bibinfo
  {author} {\bibfnamefont {A.~M.}\ \bibnamefont {Polloreno}}, \bibinfo {author}
  {\bibfnamefont {P.~J.}\ \bibnamefont {Karalekas}}, \bibinfo {author}
  {\bibfnamefont {S.}~\bibnamefont {Heidel}}, \bibinfo {author} {\bibfnamefont
  {M.~S.}\ \bibnamefont {Alam}}, \bibinfo {author} {\bibfnamefont {G.~E.}\
  \bibnamefont {Crooks}}, \ and\ \bibinfo {author} {\bibfnamefont {M.~P.}\
  \bibnamefont {da~Silva}},\ }\href {https://arxiv.org/abs/1806.08321}
  {\enquote {\bibinfo {title} {Quantum kitchen sinks: An algorithm for machine
  learning on near-term quantum computers},}\ } (\bibinfo {year} {2019}),\
  \Eprint {http://arxiv.org/abs/1806.08321} {arXiv:1806.08321 [quant-ph]}
  \BibitemShut {NoStop}%
\bibitem [{\citenamefont {Huang}\ and\ \citenamefont
  {Rebentrost}(2024)}]{PostVariationalQNNs}%
  \BibitemOpen
  \bibfield  {author} {\bibinfo {author} {\bibfnamefont {P.-W.}\ \bibnamefont
  {Huang}}\ and\ \bibinfo {author} {\bibfnamefont {P.}~\bibnamefont
  {Rebentrost}},\ }\href {https://arxiv.org/abs/2307.10560} {\enquote {\bibinfo
  {title} {Post-variational quantum neural networks},}\ } (\bibinfo {year}
  {2024}),\ \Eprint {http://arxiv.org/abs/2307.10560} {arXiv:2307.10560
  [quant-ph]} \BibitemShut {NoStop}%
\bibitem [{\citenamefont {Farhi}\ and\ \citenamefont
  {Neven}(2018)}]{GoogleQuantumMNIST}%
  \BibitemOpen
  \bibfield  {author} {\bibinfo {author} {\bibfnamefont {E.}~\bibnamefont
  {Farhi}}\ and\ \bibinfo {author} {\bibfnamefont {H.}~\bibnamefont {Neven}},\
  }\href {https://arxiv.org/abs/1802.06002} {\enquote {\bibinfo {title}
  {Classification with quantum neural networks on near term processors},}\ }
  (\bibinfo {year} {2018}),\ \Eprint {http://arxiv.org/abs/1802.06002}
  {arXiv:1802.06002 [quant-ph]} \BibitemShut {NoStop}%
\bibitem [{\citenamefont {Cong}\ \emph {et~al.}(2019)\citenamefont {Cong},
  \citenamefont {Choi},\ and\ \citenamefont {Lukin}}]{qCNN}%
  \BibitemOpen
  \bibfield  {author} {\bibinfo {author} {\bibfnamefont {I.}~\bibnamefont
  {Cong}}, \bibinfo {author} {\bibfnamefont {S.}~\bibnamefont {Choi}}, \ and\
  \bibinfo {author} {\bibfnamefont {M.~D.}\ \bibnamefont {Lukin}},\ }\href
  {\doibase 10.1038/s41567-019-0648-8} {\bibfield  {journal} {\bibinfo
  {journal} {Nature Physics}\ }\textbf {\bibinfo {volume} {15}},\ \bibinfo
  {pages} {1273–1278} (\bibinfo {year} {2019})}\BibitemShut {NoStop}%
\bibitem [{\citenamefont {Oh}\ \emph {et~al.}(2020)\citenamefont {Oh},
  \citenamefont {Choi},\ and\ \citenamefont {Kim}}]{qCNN2}%
  \BibitemOpen
  \bibfield  {author} {\bibinfo {author} {\bibfnamefont {S.}~\bibnamefont
  {Oh}}, \bibinfo {author} {\bibfnamefont {J.}~\bibnamefont {Choi}}, \ and\
  \bibinfo {author} {\bibfnamefont {J.}~\bibnamefont {Kim}},\ }\href
  {https://arxiv.org/abs/2009.09423} {\enquote {\bibinfo {title} {A tutorial on
  quantum convolutional neural networks (qcnn)},}\ } (\bibinfo {year} {2020}),\
  \Eprint {http://arxiv.org/abs/2009.09423} {arXiv:2009.09423 [quant-ph]}
  \BibitemShut {NoStop}%
\bibitem [{\citenamefont {Latorre}(2005)}]{latorre2005image}%
  \BibitemOpen
  \bibfield  {author} {\bibinfo {author} {\bibfnamefont {J.~I.}\ \bibnamefont
  {Latorre}},\ }\href@noop {} {\bibfield  {journal} {\bibinfo  {journal} {arXiv
  preprint quant-ph/0510031}\ } (\bibinfo {year} {2005})}\BibitemShut {NoStop}%
\bibitem [{\citenamefont {Cramer}\ \emph {et~al.}(2010)\citenamefont {Cramer},
  \citenamefont {Plenio}, \citenamefont {Flammia}, \citenamefont {Somma},
  \citenamefont {Gross}, \citenamefont {Bartlett}, \citenamefont
  {Landon-Cardinal}, \citenamefont {Poulin},\ and\ \citenamefont
  {Liu}}]{cramer2010efficient}%
  \BibitemOpen
  \bibfield  {author} {\bibinfo {author} {\bibfnamefont {M.}~\bibnamefont
  {Cramer}}, \bibinfo {author} {\bibfnamefont {M.~B.}\ \bibnamefont {Plenio}},
  \bibinfo {author} {\bibfnamefont {S.~T.}\ \bibnamefont {Flammia}}, \bibinfo
  {author} {\bibfnamefont {R.}~\bibnamefont {Somma}}, \bibinfo {author}
  {\bibfnamefont {D.}~\bibnamefont {Gross}}, \bibinfo {author} {\bibfnamefont
  {S.~D.}\ \bibnamefont {Bartlett}}, \bibinfo {author} {\bibfnamefont
  {O.}~\bibnamefont {Landon-Cardinal}}, \bibinfo {author} {\bibfnamefont
  {D.}~\bibnamefont {Poulin}}, \ and\ \bibinfo {author} {\bibfnamefont {Y.-K.}\
  \bibnamefont {Liu}},\ }\href@noop {} {\bibfield  {journal} {\bibinfo
  {journal} {Nature communications}\ }\textbf {\bibinfo {volume} {1}},\
  \bibinfo {pages} {149} (\bibinfo {year} {2010})}\BibitemShut {NoStop}%
\bibitem [{\citenamefont {Cai}\ \emph {et~al.}(2024)\citenamefont {Cai},
  \citenamefont {Jiang}, \citenamefont {Wang}, \citenamefont {Tang},
  \citenamefont {Kim},\ and\ \citenamefont {Huang}}]{MoEreview}%
  \BibitemOpen
  \bibfield  {author} {\bibinfo {author} {\bibfnamefont {W.}~\bibnamefont
  {Cai}}, \bibinfo {author} {\bibfnamefont {J.}~\bibnamefont {Jiang}}, \bibinfo
  {author} {\bibfnamefont {F.}~\bibnamefont {Wang}}, \bibinfo {author}
  {\bibfnamefont {J.}~\bibnamefont {Tang}}, \bibinfo {author} {\bibfnamefont
  {S.}~\bibnamefont {Kim}}, \ and\ \bibinfo {author} {\bibfnamefont
  {J.}~\bibnamefont {Huang}},\ }\href {https://arxiv.org/abs/2407.06204}
  {\enquote {\bibinfo {title} {A survey on mixture of experts},}\ } (\bibinfo
  {year} {2024}),\ \Eprint {http://arxiv.org/abs/2407.06204} {arXiv:2407.06204
  [cs.LG]} \BibitemShut {NoStop}%
\bibitem [{\citenamefont {Liu}\ \emph {et~al.}(2024)\citenamefont {Liu},
  \citenamefont {Feng}, \citenamefont {Xue}, \citenamefont {Wang},
  \citenamefont {Wu}, \citenamefont {Lu}, \citenamefont {Zhao}, \citenamefont
  {Deng}, \citenamefont {Zhang}, \citenamefont {Ruan} \emph
  {et~al.}}]{liu2024deepseek}%
  \BibitemOpen
  \bibfield  {author} {\bibinfo {author} {\bibfnamefont {A.}~\bibnamefont
  {Liu}}, \bibinfo {author} {\bibfnamefont {B.}~\bibnamefont {Feng}}, \bibinfo
  {author} {\bibfnamefont {B.}~\bibnamefont {Xue}}, \bibinfo {author}
  {\bibfnamefont {B.}~\bibnamefont {Wang}}, \bibinfo {author} {\bibfnamefont
  {B.}~\bibnamefont {Wu}}, \bibinfo {author} {\bibfnamefont {C.}~\bibnamefont
  {Lu}}, \bibinfo {author} {\bibfnamefont {C.}~\bibnamefont {Zhao}}, \bibinfo
  {author} {\bibfnamefont {C.}~\bibnamefont {Deng}}, \bibinfo {author}
  {\bibfnamefont {C.}~\bibnamefont {Zhang}}, \bibinfo {author} {\bibfnamefont
  {C.}~\bibnamefont {Ruan}},  \emph {et~al.},\ }\href@noop {} {\bibfield
  {journal} {\bibinfo  {journal} {arXiv preprint arXiv:2412.19437}\ } (\bibinfo
  {year} {2024})}\BibitemShut {NoStop}%
\bibitem [{\citenamefont {{Pennylane.ai}}(2025)}]{ParameterShiftRule}%
  \BibitemOpen
  \bibfield  {author} {\bibinfo {author} {\bibnamefont {{Pennylane.ai}}},\
  }\href {https://pennylane.ai/qml/glossary/parameter_shift} {\enquote
  {\bibinfo {title} {{Parameter-shift Rule}},}\ } (\bibinfo {year}
  {2025})\BibitemShut {NoStop}%
\bibitem [{\citenamefont {Girardi}\ and\ \citenamefont
  {De~Palma}(2025)}]{Girardi_2025}%
  \BibitemOpen
  \bibfield  {author} {\bibinfo {author} {\bibfnamefont {F.}~\bibnamefont
  {Girardi}}\ and\ \bibinfo {author} {\bibfnamefont {G.}~\bibnamefont
  {De~Palma}},\ }\href {\doibase 10.1007/s00220-025-05238-0} {\bibfield
  {journal} {\bibinfo  {journal} {Communications in Mathematical Physics}\
  }\textbf {\bibinfo {volume} {406}} (\bibinfo {year} {2025}),\
  10.1007/s00220-025-05238-0}\BibitemShut {NoStop}%
\bibitem [{\citenamefont {Hernandez}\ \emph {et~al.}(2024)\citenamefont
  {Hernandez}, \citenamefont {Girardi}, \citenamefont {Pastorello},\ and\
  \citenamefont {Palma}}]{hernandez2024quantitativeconvergencetrainedquantum}%
  \BibitemOpen
  \bibfield  {author} {\bibinfo {author} {\bibfnamefont {A.~M.}\ \bibnamefont
  {Hernandez}}, \bibinfo {author} {\bibfnamefont {F.}~\bibnamefont {Girardi}},
  \bibinfo {author} {\bibfnamefont {D.}~\bibnamefont {Pastorello}}, \ and\
  \bibinfo {author} {\bibfnamefont {G.~D.}\ \bibnamefont {Palma}},\ }\href
  {https://arxiv.org/abs/2412.03182} {\enquote {\bibinfo {title} {Quantitative
  convergence of trained quantum neural networks to a gaussian process},}\ }
  (\bibinfo {year} {2024}),\ \Eprint {http://arxiv.org/abs/2412.03182}
  {arXiv:2412.03182 [quant-ph]} \BibitemShut {NoStop}%
\bibitem [{\citenamefont {Abedi}\ \emph {et~al.}(2023)\citenamefont {Abedi},
  \citenamefont {Beigi},\ and\ \citenamefont {Taghavi}}]{Abedi_2023}%
  \BibitemOpen
  \bibfield  {author} {\bibinfo {author} {\bibfnamefont {E.}~\bibnamefont
  {Abedi}}, \bibinfo {author} {\bibfnamefont {S.}~\bibnamefont {Beigi}}, \ and\
  \bibinfo {author} {\bibfnamefont {L.}~\bibnamefont {Taghavi}},\ }\href
  {\doibase 10.22331/q-2023-04-27-989} {\bibfield  {journal} {\bibinfo
  {journal} {Quantum}\ }\textbf {\bibinfo {volume} {7}},\ \bibinfo {pages}
  {989} (\bibinfo {year} {2023})}\BibitemShut {NoStop}%
\bibitem [{\citenamefont {Bergholm}\ \emph {et~al.}(2022)\citenamefont
  {Bergholm}, \citenamefont {Izaac}, \citenamefont {Schuld}, \citenamefont
  {Gogolin}, \citenamefont {Ahmed}, \citenamefont {Ajith}, \citenamefont
  {Alam}, \citenamefont {Alonso-Linaje}, \citenamefont {AkashNarayanan},
  \citenamefont {Asadi}, \citenamefont {Arrazola}, \citenamefont {Azad},
  \citenamefont {Banning}, \citenamefont {Blank}, \citenamefont {Bromley},
  \citenamefont {Cordier}, \citenamefont {Ceroni}, \citenamefont {Delgado},
  \citenamefont {Matteo}, \citenamefont {Dusko}, \citenamefont {Garg},
  \citenamefont {Guala}, \citenamefont {Hayes}, \citenamefont {Hill},
  \citenamefont {Ijaz}, \citenamefont {Isacsson}, \citenamefont {Ittah},
  \citenamefont {Jahangiri}, \citenamefont {Jain}, \citenamefont {Jiang},
  \citenamefont {Khandelwal}, \citenamefont {Kottmann}, \citenamefont {Lang},
  \citenamefont {Lee}, \citenamefont {Loke}, \citenamefont {Lowe},
  \citenamefont {McKiernan}, \citenamefont {Meyer}, \citenamefont
  {Montañez-Barrera}, \citenamefont {Moyard}, \citenamefont {Niu},
  \citenamefont {O'Riordan}, \citenamefont {Oud}, \citenamefont {Panigrahi},
  \citenamefont {Park}, \citenamefont {Polatajko}, \citenamefont {Quesada},
  \citenamefont {Roberts}, \citenamefont {Sá}, \citenamefont {Schoch},
  \citenamefont {Shi}, \citenamefont {Shu}, \citenamefont {Sim}, \citenamefont
  {Singh}, \citenamefont {Strandberg}, \citenamefont {Soni}, \citenamefont
  {Száva}, \citenamefont {Thabet}, \citenamefont {Vargas-Hernández},
  \citenamefont {Vincent}, \citenamefont {Vitucci}, \citenamefont {Weber},
  \citenamefont {Wierichs}, \citenamefont {Wiersema}, \citenamefont {Willmann},
  \citenamefont {Wong}, \citenamefont {Zhang},\ and\ \citenamefont
  {Killoran}}]{bergholm2018pennylane}%
  \BibitemOpen
  \bibfield  {author} {\bibinfo {author} {\bibfnamefont {V.}~\bibnamefont
  {Bergholm}}, \bibinfo {author} {\bibfnamefont {J.}~\bibnamefont {Izaac}},
  \bibinfo {author} {\bibfnamefont {M.}~\bibnamefont {Schuld}}, \bibinfo
  {author} {\bibfnamefont {C.}~\bibnamefont {Gogolin}}, \bibinfo {author}
  {\bibfnamefont {S.}~\bibnamefont {Ahmed}}, \bibinfo {author} {\bibfnamefont
  {V.}~\bibnamefont {Ajith}}, \bibinfo {author} {\bibfnamefont {M.~S.}\
  \bibnamefont {Alam}}, \bibinfo {author} {\bibfnamefont {G.}~\bibnamefont
  {Alonso-Linaje}}, \bibinfo {author} {\bibfnamefont {B.}~\bibnamefont
  {AkashNarayanan}}, \bibinfo {author} {\bibfnamefont {A.}~\bibnamefont
  {Asadi}}, \bibinfo {author} {\bibfnamefont {J.~M.}\ \bibnamefont {Arrazola}},
  \bibinfo {author} {\bibfnamefont {U.}~\bibnamefont {Azad}}, \bibinfo {author}
  {\bibfnamefont {S.}~\bibnamefont {Banning}}, \bibinfo {author} {\bibfnamefont
  {C.}~\bibnamefont {Blank}}, \bibinfo {author} {\bibfnamefont {T.~R.}\
  \bibnamefont {Bromley}}, \bibinfo {author} {\bibfnamefont {B.~A.}\
  \bibnamefont {Cordier}}, \bibinfo {author} {\bibfnamefont {J.}~\bibnamefont
  {Ceroni}}, \bibinfo {author} {\bibfnamefont {A.}~\bibnamefont {Delgado}},
  \bibinfo {author} {\bibfnamefont {O.~D.}\ \bibnamefont {Matteo}}, \bibinfo
  {author} {\bibfnamefont {A.}~\bibnamefont {Dusko}}, \bibinfo {author}
  {\bibfnamefont {T.}~\bibnamefont {Garg}}, \bibinfo {author} {\bibfnamefont
  {D.}~\bibnamefont {Guala}}, \bibinfo {author} {\bibfnamefont
  {A.}~\bibnamefont {Hayes}}, \bibinfo {author} {\bibfnamefont
  {R.}~\bibnamefont {Hill}}, \bibinfo {author} {\bibfnamefont {A.}~\bibnamefont
  {Ijaz}}, \bibinfo {author} {\bibfnamefont {T.}~\bibnamefont {Isacsson}},
  \bibinfo {author} {\bibfnamefont {D.}~\bibnamefont {Ittah}}, \bibinfo
  {author} {\bibfnamefont {S.}~\bibnamefont {Jahangiri}}, \bibinfo {author}
  {\bibfnamefont {P.}~\bibnamefont {Jain}}, \bibinfo {author} {\bibfnamefont
  {E.}~\bibnamefont {Jiang}}, \bibinfo {author} {\bibfnamefont
  {A.}~\bibnamefont {Khandelwal}}, \bibinfo {author} {\bibfnamefont
  {K.}~\bibnamefont {Kottmann}}, \bibinfo {author} {\bibfnamefont {R.~A.}\
  \bibnamefont {Lang}}, \bibinfo {author} {\bibfnamefont {C.}~\bibnamefont
  {Lee}}, \bibinfo {author} {\bibfnamefont {T.}~\bibnamefont {Loke}}, \bibinfo
  {author} {\bibfnamefont {A.}~\bibnamefont {Lowe}}, \bibinfo {author}
  {\bibfnamefont {K.}~\bibnamefont {McKiernan}}, \bibinfo {author}
  {\bibfnamefont {J.~J.}\ \bibnamefont {Meyer}}, \bibinfo {author}
  {\bibfnamefont {J.~A.}\ \bibnamefont {Montañez-Barrera}}, \bibinfo {author}
  {\bibfnamefont {R.}~\bibnamefont {Moyard}}, \bibinfo {author} {\bibfnamefont
  {Z.}~\bibnamefont {Niu}}, \bibinfo {author} {\bibfnamefont {L.~J.}\
  \bibnamefont {O'Riordan}}, \bibinfo {author} {\bibfnamefont {S.}~\bibnamefont
  {Oud}}, \bibinfo {author} {\bibfnamefont {A.}~\bibnamefont {Panigrahi}},
  \bibinfo {author} {\bibfnamefont {C.-Y.}\ \bibnamefont {Park}}, \bibinfo
  {author} {\bibfnamefont {D.}~\bibnamefont {Polatajko}}, \bibinfo {author}
  {\bibfnamefont {N.}~\bibnamefont {Quesada}}, \bibinfo {author} {\bibfnamefont
  {C.}~\bibnamefont {Roberts}}, \bibinfo {author} {\bibfnamefont
  {N.}~\bibnamefont {Sá}}, \bibinfo {author} {\bibfnamefont {I.}~\bibnamefont
  {Schoch}}, \bibinfo {author} {\bibfnamefont {B.}~\bibnamefont {Shi}},
  \bibinfo {author} {\bibfnamefont {S.}~\bibnamefont {Shu}}, \bibinfo {author}
  {\bibfnamefont {S.}~\bibnamefont {Sim}}, \bibinfo {author} {\bibfnamefont
  {A.}~\bibnamefont {Singh}}, \bibinfo {author} {\bibfnamefont
  {I.}~\bibnamefont {Strandberg}}, \bibinfo {author} {\bibfnamefont
  {J.}~\bibnamefont {Soni}}, \bibinfo {author} {\bibfnamefont {A.}~\bibnamefont
  {Száva}}, \bibinfo {author} {\bibfnamefont {S.}~\bibnamefont {Thabet}},
  \bibinfo {author} {\bibfnamefont {R.~A.}\ \bibnamefont {Vargas-Hernández}},
  \bibinfo {author} {\bibfnamefont {T.}~\bibnamefont {Vincent}}, \bibinfo
  {author} {\bibfnamefont {N.}~\bibnamefont {Vitucci}}, \bibinfo {author}
  {\bibfnamefont {M.}~\bibnamefont {Weber}}, \bibinfo {author} {\bibfnamefont
  {D.}~\bibnamefont {Wierichs}}, \bibinfo {author} {\bibfnamefont
  {R.}~\bibnamefont {Wiersema}}, \bibinfo {author} {\bibfnamefont
  {M.}~\bibnamefont {Willmann}}, \bibinfo {author} {\bibfnamefont
  {V.}~\bibnamefont {Wong}}, \bibinfo {author} {\bibfnamefont {S.}~\bibnamefont
  {Zhang}}, \ and\ \bibinfo {author} {\bibfnamefont {N.}~\bibnamefont
  {Killoran}},\ }\href {https://arxiv.org/abs/1811.04968} {\enquote {\bibinfo
  {title} {Pennylane: Automatic differentiation of hybrid quantum-classical
  computations},}\ } (\bibinfo {year} {2022}),\ \Eprint
  {http://arxiv.org/abs/1811.04968} {arXiv:1811.04968 [quant-ph]} \BibitemShut
  {NoStop}%
\bibitem [{\citenamefont {Kingma}\ and\ \citenamefont {Ba}(2017)}]{Adam}%
  \BibitemOpen
  \bibfield  {author} {\bibinfo {author} {\bibfnamefont {D.~P.}\ \bibnamefont
  {Kingma}}\ and\ \bibinfo {author} {\bibfnamefont {J.}~\bibnamefont {Ba}},\
  }\href {https://arxiv.org/abs/1412.6980} {\enquote {\bibinfo {title} {Adam: A
  method for stochastic optimization},}\ } (\bibinfo {year} {2017}),\ \Eprint
  {http://arxiv.org/abs/1412.6980} {arXiv:1412.6980 [cs.LG]} \BibitemShut
  {NoStop}%
\bibitem [{\citenamefont {P{\'e}rez-Salinas}\ \emph {et~al.}(2020)\citenamefont
  {P{\'e}rez-Salinas}, \citenamefont {Cervera-Lierta}, \citenamefont
  {Gil-Fuster},\ and\ \citenamefont {Latorre}}]{perez2020data}%
  \BibitemOpen
  \bibfield  {author} {\bibinfo {author} {\bibfnamefont {A.}~\bibnamefont
  {P{\'e}rez-Salinas}}, \bibinfo {author} {\bibfnamefont {A.}~\bibnamefont
  {Cervera-Lierta}}, \bibinfo {author} {\bibfnamefont {E.}~\bibnamefont
  {Gil-Fuster}}, \ and\ \bibinfo {author} {\bibfnamefont {J.~I.}\ \bibnamefont
  {Latorre}},\ }\href@noop {} {\bibfield  {journal} {\bibinfo  {journal}
  {Quantum}\ }\textbf {\bibinfo {volume} {4}},\ \bibinfo {pages} {226}
  (\bibinfo {year} {2020})}\BibitemShut {NoStop}%
\bibitem [{\citenamefont {Cireşan}\ \emph {et~al.}(2012)\citenamefont
  {Cireşan}, \citenamefont {Meier},\ and\ \citenamefont
  {Schmidhuber}}]{MaxAccuracyMNIST}%
  \BibitemOpen
  \bibfield  {author} {\bibinfo {author} {\bibfnamefont {D.}~\bibnamefont
  {Cireşan}}, \bibinfo {author} {\bibfnamefont {U.}~\bibnamefont {Meier}}, \
  and\ \bibinfo {author} {\bibfnamefont {J.}~\bibnamefont {Schmidhuber}},\
  }\href {https://arxiv.org/abs/1202.2745} {\bibfield  {journal} {\bibinfo
  {journal} {CVPR 2012, p. 3642-3649}\ } (\bibinfo {year} {2012})},\ \Eprint
  {http://arxiv.org/abs/1202.2745} {arXiv:1202.2745 [cs.CV]} \BibitemShut
  {NoStop}%
\bibitem [{\citenamefont {Deng}(2012)}]{MNIST}%
  \BibitemOpen
  \bibfield  {author} {\bibinfo {author} {\bibfnamefont {L.}~\bibnamefont
  {Deng}},\ }\href {\doibase 10.1109/MSP.2012.2211477} {\bibfield  {journal}
  {\bibinfo  {journal} {IEEE Signal Processing Magazine}\ }\textbf {\bibinfo
  {volume} {29}},\ \bibinfo {pages} {141} (\bibinfo {year} {2012})}\BibitemShut
  {NoStop}%
\bibitem [{\citenamefont {An}\ \emph {et~al.}(2020)\citenamefont {An},
  \citenamefont {Lee}, \citenamefont {Park}, \citenamefont {Yang},\ and\
  \citenamefont {So}}]{CNNensemble}%
  \BibitemOpen
  \bibfield  {author} {\bibinfo {author} {\bibfnamefont {S.}~\bibnamefont
  {An}}, \bibinfo {author} {\bibfnamefont {M.}~\bibnamefont {Lee}}, \bibinfo
  {author} {\bibfnamefont {S.}~\bibnamefont {Park}}, \bibinfo {author}
  {\bibfnamefont {H.}~\bibnamefont {Yang}}, \ and\ \bibinfo {author}
  {\bibfnamefont {J.}~\bibnamefont {So}},\ }\href
  {https://arxiv.org/abs/2008.10400} {\enquote {\bibinfo {title} {An ensemble
  of simple convolutional neural network models for mnist digit recognition},}\
  } (\bibinfo {year} {2020}),\ \Eprint {http://arxiv.org/abs/2008.10400}
  {arXiv:2008.10400 [cs.CV]} \BibitemShut {NoStop}%
\bibitem [{\citenamefont {Hasanpour}\ \emph {et~al.}(2023)\citenamefont
  {Hasanpour}, \citenamefont {Rouhani}, \citenamefont {Fayyaz},\ and\
  \citenamefont {Sabokrou}}]{CNNoverview}%
  \BibitemOpen
  \bibfield  {author} {\bibinfo {author} {\bibfnamefont {S.~H.}\ \bibnamefont
  {Hasanpour}}, \bibinfo {author} {\bibfnamefont {M.}~\bibnamefont {Rouhani}},
  \bibinfo {author} {\bibfnamefont {M.}~\bibnamefont {Fayyaz}}, \ and\ \bibinfo
  {author} {\bibfnamefont {M.}~\bibnamefont {Sabokrou}},\ }\href
  {https://arxiv.org/abs/1608.06037} {\enquote {\bibinfo {title} {Lets keep it
  simple, using simple architectures to outperform deeper and more complex
  architectures},}\ } (\bibinfo {year} {2023}),\ \Eprint
  {http://arxiv.org/abs/1608.06037} {arXiv:1608.06037 [cs.CV]} \BibitemShut
  {NoStop}%
\end{thebibliography}%

\clearpage

\appendix
\onecolumngrid

\section{The MNIST dataset}\label{sec:appendixMNIST}
The MNIST \cite{MNIST} is a dataset of labeled images of handwritten digits, widely used to benchmark vision models. Each image has $28 \times 28$ pixels. There are $60,000$ images in the training set and $10,000$ in the test set. We used $50,000$ images from the training set, selected with a different seed in each iteration, and we kept the remaining $10,000$ images available for potential statistical controls that eventually proved unnecessary. The $10,000$ test images are all used and are the same in each iteration.

\section{Convolutional Neural Networks}
\label{sec:appendixCNN}
\begin{figure*}[t]
	\centering
	\includegraphics[width=1.0\linewidth]{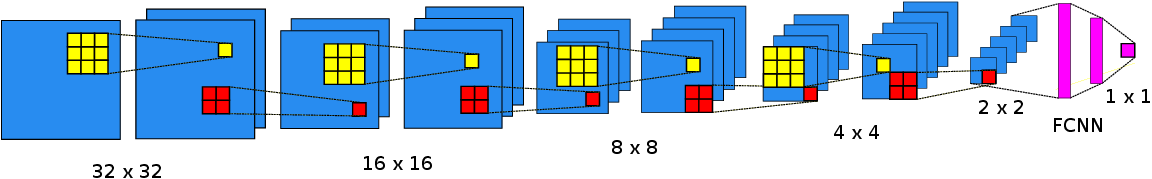}
	\caption{General structure of our classical Convolutional Neural Network. The image undergoes a series of convolutional layers with a $3 \times 3$ kernel matrix (yellow), and pooling layers (red). When a small resolution of $2\times2$ is achieved, the pixels are flattened and given to a fully connected neural network (purple). To compare the classical CNN with our quantum architecture, the number of parameters of the fully connected NN is kept small.}
	\label{fig:CNN}
\end{figure*}

\begin{figure}[t]
	\centering
	\includegraphics[width=0.5\linewidth]{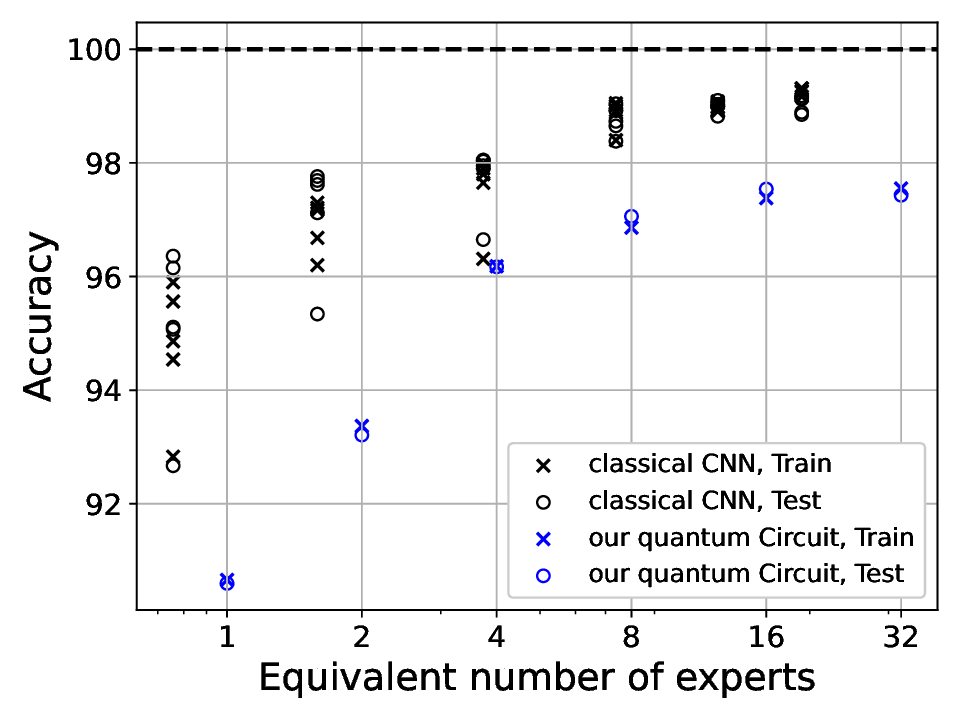}
	\caption{Comparison between the accuracy of classical and quantum neural networks in the classification of the parity of the MNIST dataset. The equivalent number of experts is simply the number of parameters of a (classical or quantum) model divided by $252$, which is the number of parameters of a quantum expert. In this way we can assign a (usually non-integer) equivalent number of expert to our classical circuits to compare them to the quantum ones.}
	\label{fig:quantum vs classical}
\end{figure}

We have compared our quantum neural network with a classical CNN with roughly the same number of parameters.
This is unusual, as the number of parameters of our quantum neural networks goes from $252$ for one expert to $8064$ for thirty-two experts, which is at least an order of magnitude smaller than the number of parameters of the typical CNNs employed for MNIST, which is more than $100,000$ and usually of the order of millions \cite{CNNensemble, CNNoverview}.\\
In this appendix we will describe how CNNs work in general, and then we will explain how we kept the number of parameters so small in our specific instance.\\
An image of dimension $n\times n$ with $c$ channels, i.e., $c$ numbers for each pixel ($c=3$ for RGB images, $c=1$ for BW images) is embedded into a tensor of size $(n, n, c)$. Since the MNIST images are black and white, our inputs have $c=1$.
\begin{figure}[t]
	\centering
	\includegraphics[width=0.45\linewidth]{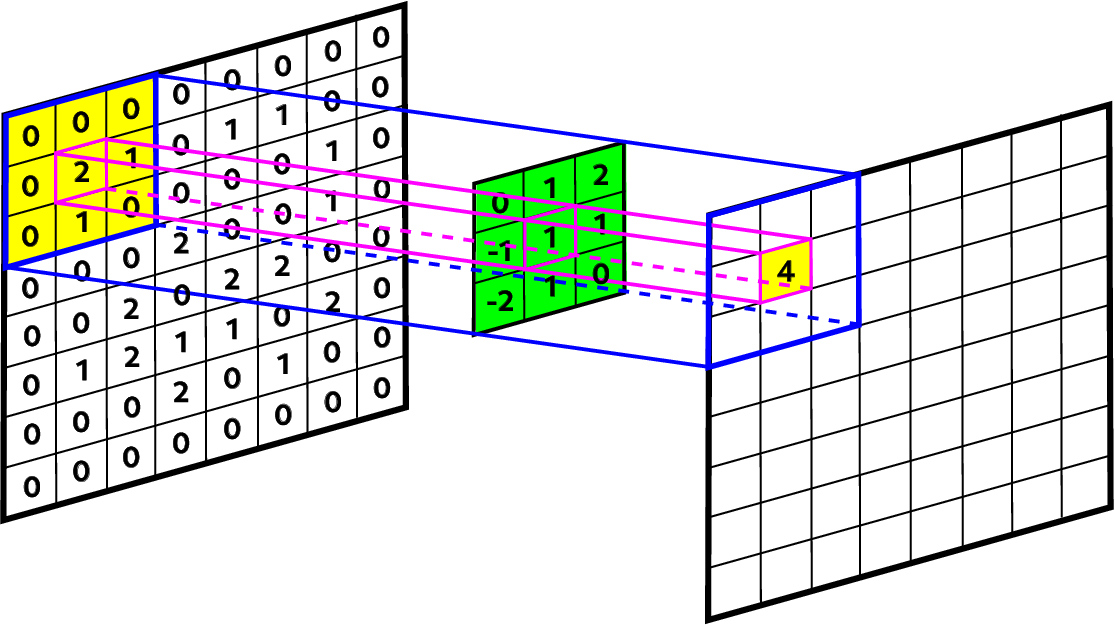}
	\caption{A convolutional layer with a $3\times3$ kernel (shown in green). The calculation that gives the result is to multiply each term in the $3\times3$ kernel (shown in green) with the corresponding term in a $3\times3$ subset of the original image (the subset is shown in yellow to the left), and then to sum the results and write it in the pixel corresponding to the central pixel of the original subset (the output pixel is shown in yellow to the right).}
	\label{fig:Convolution}
\end{figure}
In each convolutional layer, the network computes the convolution between a kernel matrix and the image (see \autoref{fig:Convolution}): using a $3\times3$ kernel, a stride of $1$ and a padding of $1$, the output of the convolutional layer will have the same dimension as the input.
If instead of a single kernel we have $c$ kernels (or analogously if the kernel is a tensor of dimension $(3,3,c)$), we will have an output image of size $(n,n,c)$.
If the starting image has already a nontrivial number of channels, e.g. dimension $(n, n, c)$, a kernel of dimension $(3, 3, c, c')$ will give, after a contraction in the indices of dimension $c$, a resulting image of dimension $(n, n, c')$. This is the most general convolutional layer that we will use.
The usual strategy is to alternate convolutional layers with pooling layers, which divide the image into blocks of size $2\times2$ and replace each block with a single pixel whose intensity is the average intensity of its original components. The pooling layers decrease the resolution of the image, preserving the information at a larger scale. If the input image has size $(n, n, c)$ with $n$ even, the output image has size $(n/2, n/2, c)$.
We embed the $28\times28$ MNIST images in the $32 \times 32$ images by adding zeroes at the boundary.
Our CNN applies five pooling layers alternated with convolutional layers.
Usually after some number of convolutional layers, the $(\bar n,\bar n,\bar c)$ output image is flattened to a vector of $\bar n^2 \bar c$ components, and then a fully connected neural network is used to calculate the output.\\
Now we focus on our version of the CNN, which we show in \autoref{fig:CNN}. We wanted a small number of parameters. As they are mostly concentrated in the final fully connected NN, we tried to use convolutions to reduce the dimensions of the input of the fully connected network as far as we could. For that reason, we applied four convolutions, interleaved with pooling layers, until our image was of dimension $2\times2$ with $c_4$ channels. Then we flattened this tensor to a $4c_4$ vector, and this is the input of the final fully connected NN with hidden layer $h$ and output layer $1$. The output is a single number, the sign of which gives the prediction of our CNN binary classifier.\\
As classical CNNs have almost all parameters concentrated in the fully connected neural network at the end, we pay particular attention to keeping $c_4$ and $h$ low. The parameters used by our classical CNNs are shown in \autoref{tab:table2}.
\begin{table*}[t]
	\centering
	\begin{tabular}{l c c c c c c c r}
	    \toprule
	    \#q par.s & \#cl par.s & $c_1$ & $c_2$ & $c_3$ & $c_4$ & $h$ & train acc. & test acc.\\
	    \midrule
	    $252$ & $175$ & $2$ & $2$ & $2$ & $4$ & $4$ & $94.74 \pm 1.14$ & $95.07 \pm 1.72$\\
	    $504$ & $465$ & $2$ & $4$ & $4$ & $4$ & $4$ & $96.91 \pm 0.17$ & $97.11 \pm 0.83$\\
	    $1008$ & $905$ & $4$ & $4$ & $4$ & $8$ & $8$ & $97.53 \pm 0.38$ & $97.72 \pm 0.29$\\
	    $2016$ & $1969$ & $8$ & $8$ & $8$ & $8$ & $4$ & $98.85 \pm 0.05$ & $98.74 \pm 0.05$\\
	    $4032$ & $3969$ & $16$ & $16$ & $8$ & $4$ & $2$ & $99.000 \pm 0.004$ & $98.99 \pm 0.01$\\
	    $8064$ & $7829$ & $32$ & $16$ & $12$ & $8$ & $8$ & $99.262 \pm 0.003$ & $99.04 \pm 0.02$\\
	    \bottomrule
	\end{tabular}
    \caption{Parameters of the classical CNNs: $c_1,\ c_2,\ c_3,\ c_4$ are the number of channels of the output of the four convolutional layers respectively, while $h$ is the number of neurons of the hidden layer of the fully connected NN at the last step of the CNN. We present train and test accuracy, and the number of parameters next to that of the quantum circuit we want to compare it to.}
    \label{tab:table2}
\end{table*}
\autoref{fig:quantum vs classical} shows that our quantum neural networks can achieve a worse test accuracy than the classical CNNs with a similar number of parameters, which is probably due to the more general non-linearity of the classical CNN, while our quantum circuit accuracy is upper bounded by the quadratic classifier accuracy, as shown in Appendix~\ref{sec:AppendixQuadratic}.

\section{Convolutions with a single 1-qubit gate}\label{sec:1 qubit convolution}
We argue that, due to amplitude encoding, a $1$-qubit gate in our circuit corresponds to a classical convolution operation. Let us imagine for simplicity an operation on $y_4$, the higher resolution pixel on the $y$ coordinate of the image (see \autoref{fig:gates}). If we apply to that qubit a $1$-qubit gate that rotates $\ket{0}$ and $\ket{1}$, then we are mixing all pixels of the image whose $y$ coordinate ends in $0$ with the nearby pixel where it ends in $1$. This is in fact a convolution with a $1\times2$ kernel, whose result is overwritten to the original image. To see it, let us take the general matrix
\[\begin{pmatrix}
\alpha & \beta \\
\gamma & \delta
\end{pmatrix}\]
that acts mixing $\ket{s0}$ and $\ket{s1}$ for all bit strings $s$. This means that the amplitudes $c_{s0}$ and $c_{s1}$ become:
\[c_{s0}' = \alpha c_{s0} + \beta c_{s1}\]
\[c_{s1}' = \gamma c_{s0} + \delta c_{s1}\]
For a real matrix this corresponds to a convolution by the $1\times2$ matrix $\begin{pmatrix}\alpha & \beta \end{pmatrix}$ which updates the odd $y$ pixels acting with horizontal stride $2$, and $\begin{pmatrix}\gamma & \delta \end{pmatrix}$ which updates on the even $y$ pixels acting with horizontal stride $2$ - the stride means simply that we are applying the operation on every other pixel in the horizontal direction.\\
Classically, this operation would depend on few parameters but would have to be done repeatedly throughout the image. In our quantum circuit, this operation depends on a single gate that, thanks to amplitude encoding, can be applied just once.

\section{Extended numerical results}\label{sec:full}
We present here all the data obtained in the simulations. 
In each epoch, we have computed the \textbf{running train accuracy}, i.e.~the percentage of right predictions on the training images of the selected batches, computed at ``run-time'' during training when the parameters are evolving. 
The train accuracy (with the final parameters) is calculated only at the end of training.
The \textbf{test accuracy} is calculated at the end of each epoch with fixed parameters for all test images.
The number of epochs is chosen to saturate the accuracy and varies with the number of experts.

\begin{figure*}[htpb]
  \begin{center}
    \begin{tikzpicture}[scale=1, transform shape]
      \node at (0,0) {
          \includegraphics[width=0.48\linewidth]{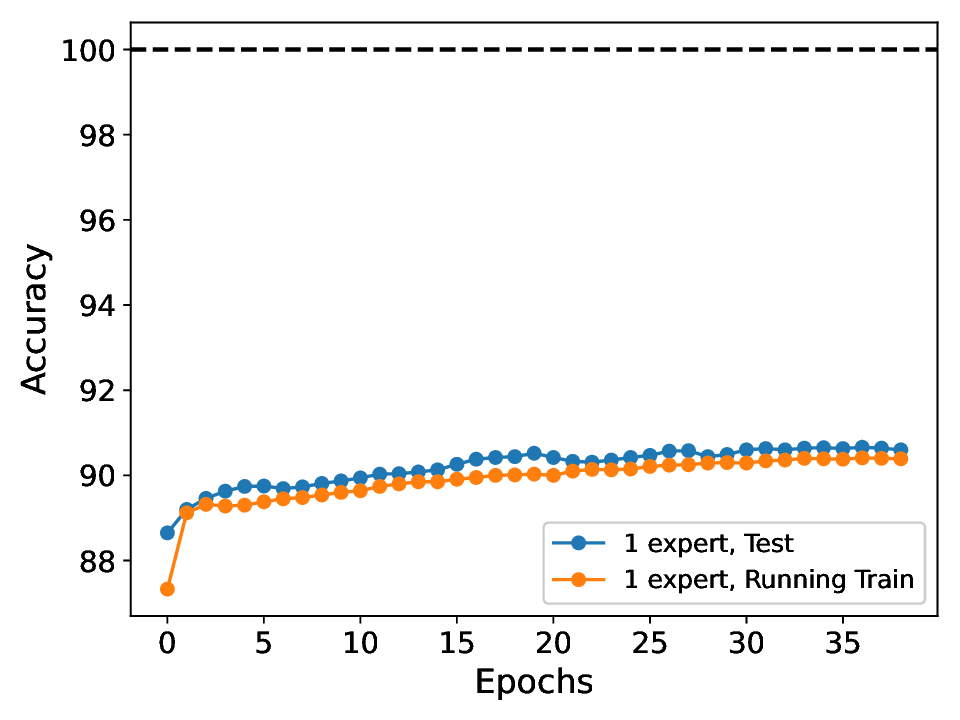}
        }; 
      \node at (9,0) {
          \includegraphics[width=0.48\linewidth]{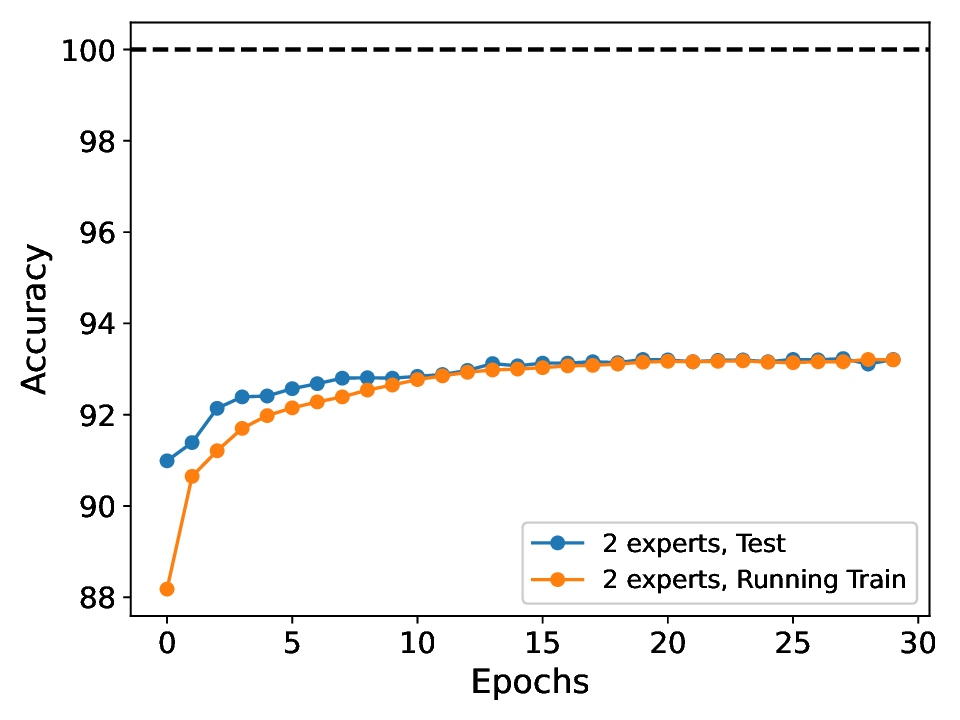}
        };
      \node at (0,-6.5) {
          \includegraphics[width=0.48\linewidth]{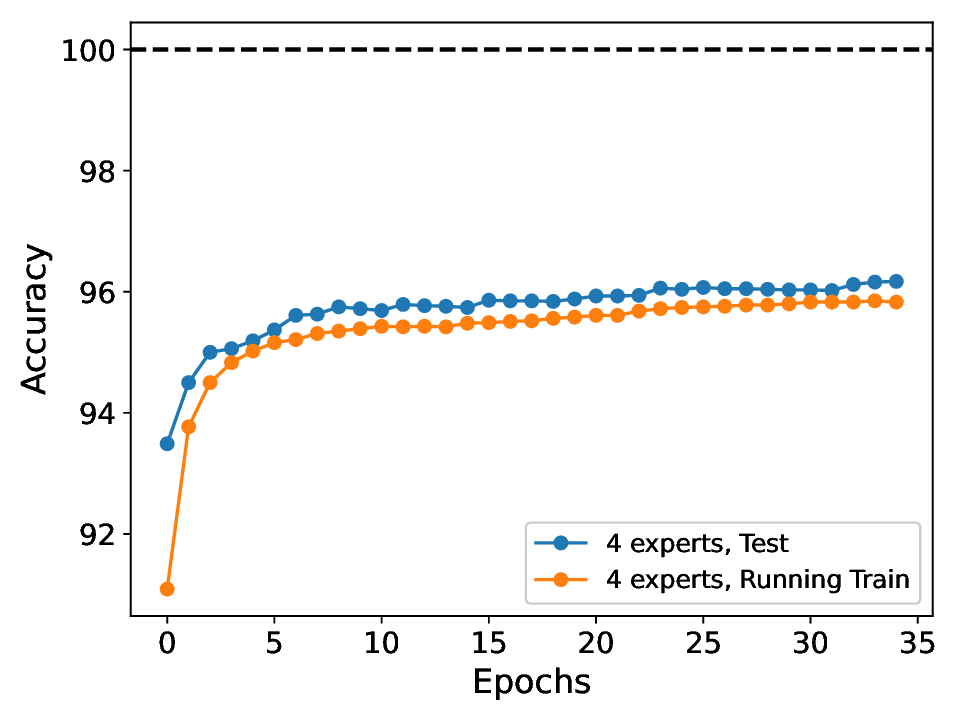}
        };
      \node at (9,-6.5) {
          \includegraphics[width=0.48\linewidth]{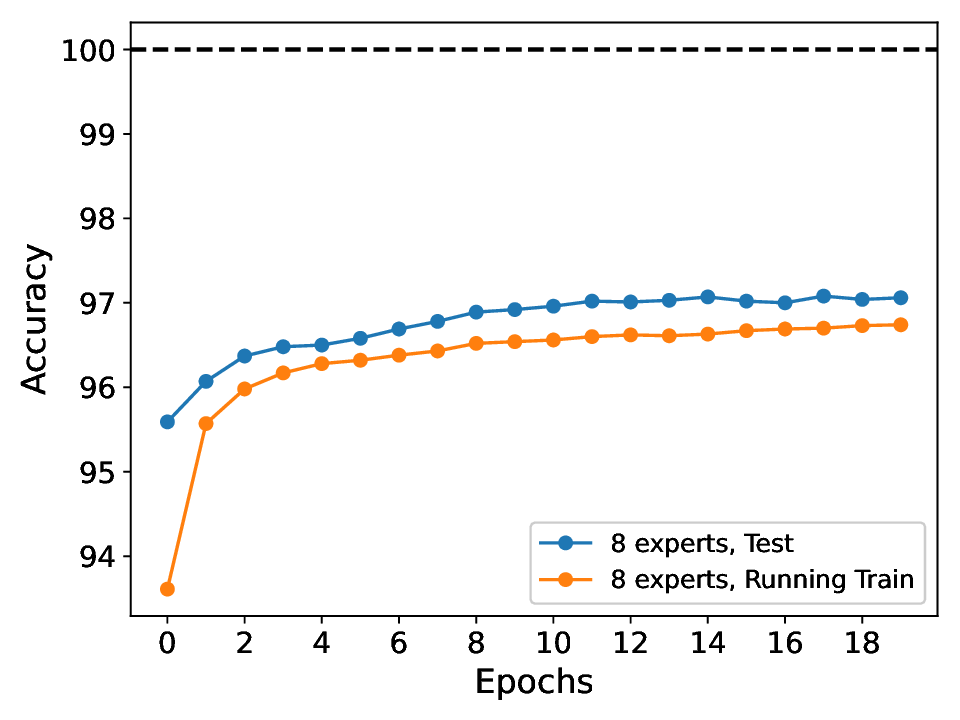}
        };
      \node at (0,-13) {
          \includegraphics[width=0.48\linewidth]{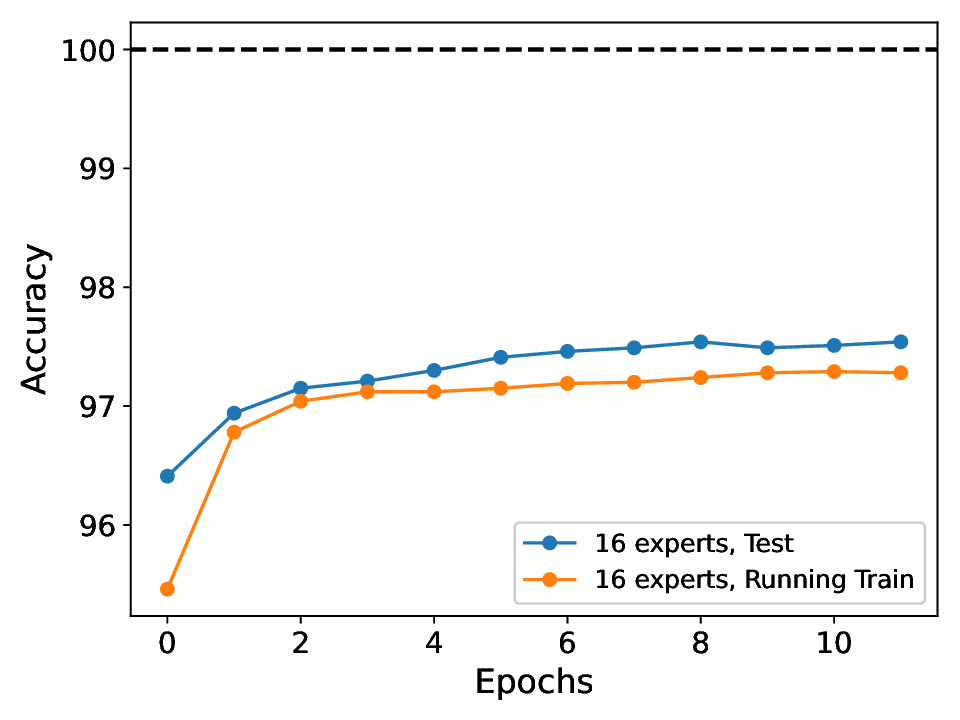}
        };
      \node at (9,-13) {
          \includegraphics[width=0.48\linewidth]{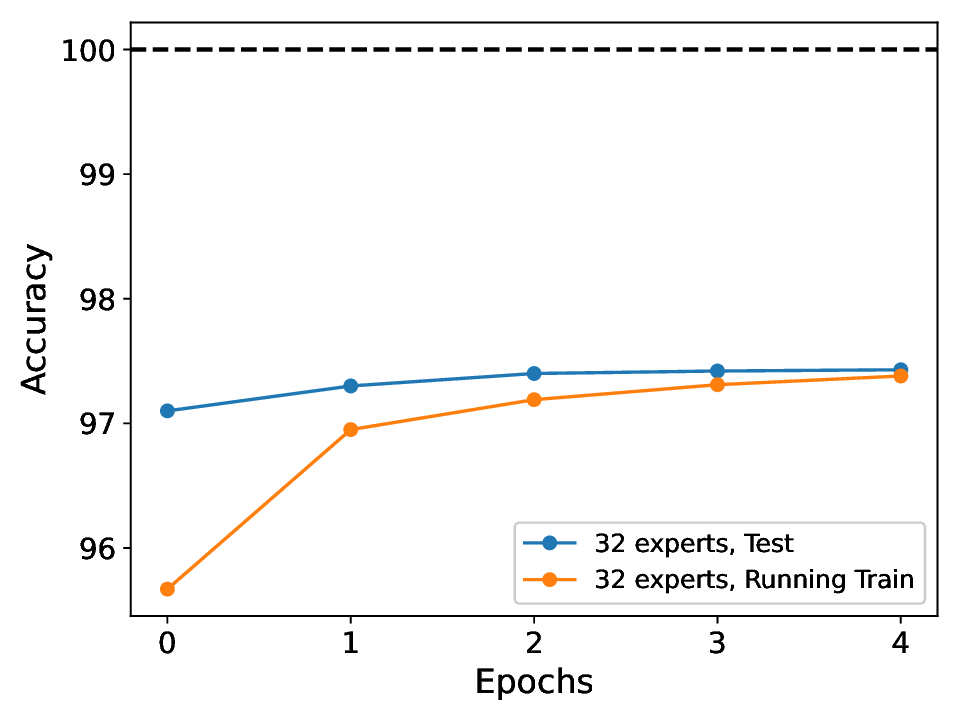}
        };
    \end{tikzpicture}
  \end{center}
  \caption{Running and Test Accuracy using circuits with different experts. }
  \label{fig:extended results}
\end{figure*}

First we present a graph for each of the quantum neural networks we simulated (see \autoref{fig:extended results}), each with a different number of experts, showing the running train accuracy and the test accuracy.
We notice that the test accuracy is greater than the running train accuracy. This is due to the fact that the two accuracies are not computed with the same parameters, since the running train accuracy is computed during the training while the parameters of the model are changing, while the test accuracy is computed with fixed parameters at the end of each epoch.
In fact, we have computed the train accuracy with the parameters at the end of the training, and it turns out to be higher or comparable to the test accuracy as expected (see \autoref{tab:table1}).

\begin{figure*}[t]
	\centering
	\includegraphics[width=0.45\linewidth]{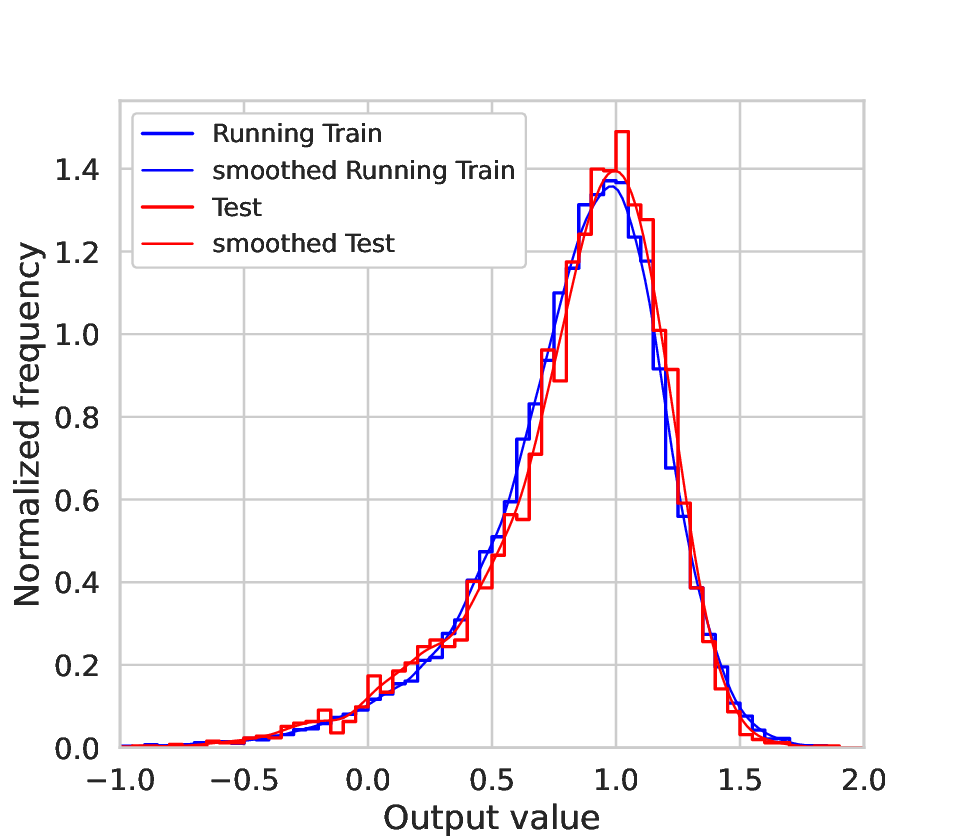}
    \includegraphics[width=0.45\linewidth]{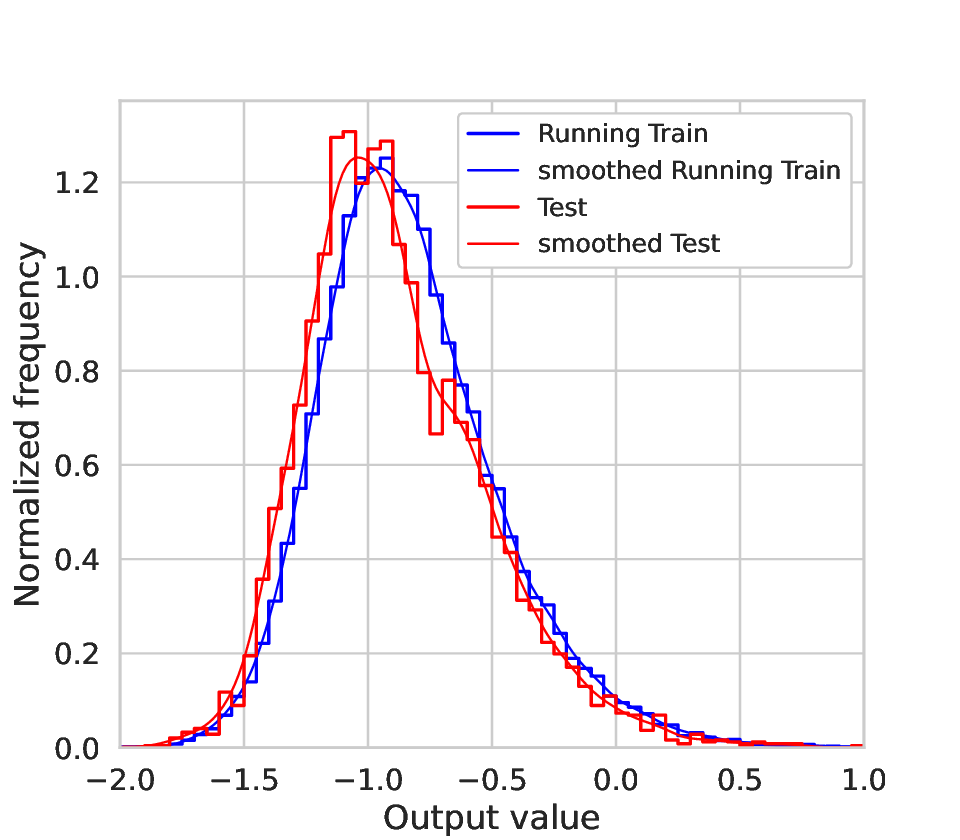}
	\caption{Histogram of circuit outputs for odd (left) and even (right) images.}
	\label{fig:Histogram}
\end{figure*}

We also show the histograms that represent the distributions of the output of the circuit, given the different train or test images of the dataset. We show it only for the trained circuit with 16 experts. As expected, the outputs of the trained circuit concentrate around $+1$ and $-1$ for odd and even images, respectively (see \autoref{fig:Histogram}).

\begin{figure}[t]
	\centering
	\includegraphics[width=0.5\linewidth]{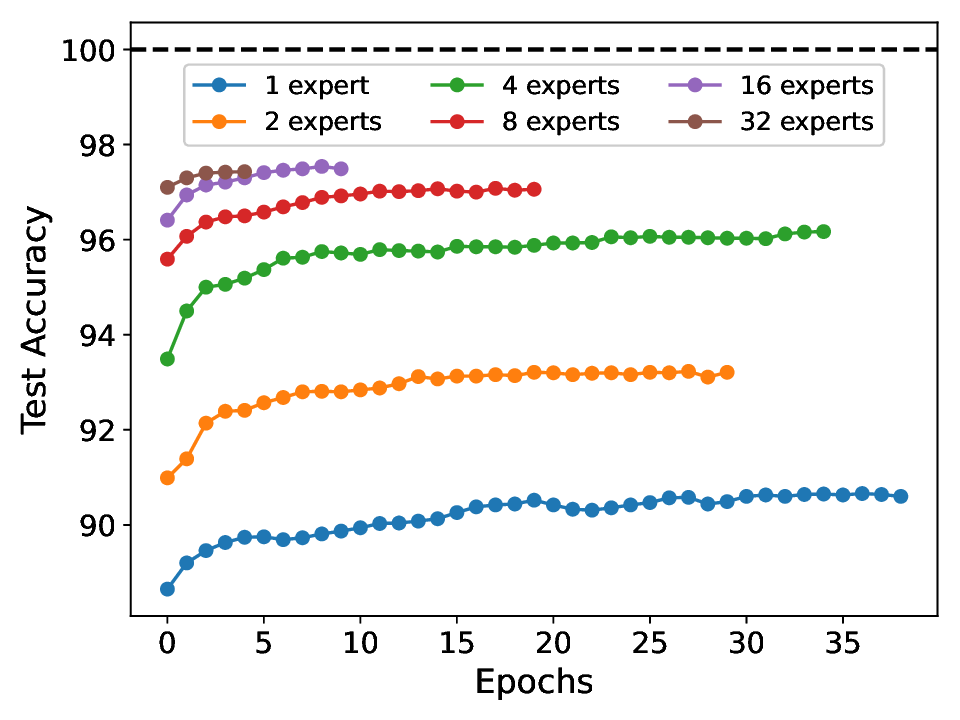}
	\caption{Test accuracy in function of the epochs. Here we presented the complete data.}
	\label{fig:test vs epochs}
\end{figure}

The test accuracy of all simulations as a function of the epoch is shown in \autoref{fig:test vs epochs}, using the same data of \autoref{fig:test vs epochs rescaled}.

\section{Classical Quadratic classifier}
\label{sec:AppendixQuadratic}
We notice that the function generated by each expert is quadratic in the wavefunction of the input of the expert, which is the vector of the intensities of the pixels (except for the normalization factor).
Therefore, the model function of our quantum neural network is a quadratic polynomial of the vector of the intensities of the pixels.
Thus, we compare our results with a \textbf{classical quadratic classifier}, which for $784$ pixels has $1 + 784 + \binom{784+1}{2} = 308,505$ parameters, and
achieves a test accuracy of about $98\%$, which is right above the maximum accuracy reached by our model ($97.5\%$).

\end{document}